\documentclass[twocolumn]{aastex62}
\usepackage{bm}
\usepackage{amsmath}

\newcommand\St{\mathrm{St}}
\newcommand\Rey{\mathrm{Re}}
\newcommand{\tatsuuma}{} 
\newcommand{\misako}{} 
\newcommand{\mtatsuuma}{} 
\newcommand{\fourthcomment}{} 
\newcommand{\fifthcomment}{} 

\graphicspath{{./}{figures/}}

\received{December 10, 2019}
\revised{March 18, 2021}
\accepted{April 6, 2021}
\submitjournal{\apj}

\shorttitle{Rotational Disruption of Dust Aggregates} 
\shortauthors{Tatsuuma \& Kataoka}

\begin{document}

\title{Rotational Disruption of Porous Dust Aggregates due to Gas Flow in Protoplanetary Disks}

\correspondingauthor{Misako Tatsuuma}
\email{misako.tatsuuma@gmail.com}

\author[0000-0003-1844-5107]{Misako Tatsuuma}
\affiliation{Department of Astronomy, Graduate School of Science,
The University of Tokyo, 7-3-1 Hongo, Bunkyo-ku, Tokyo 113-0033, Japan}
\affiliation{Division of Science, National Astronomical Observatory of Japan,
2-21-1 Osawa, Mitaka, Tokyo 181-8588, Japan}

\author[0000-0003-4562-4119]{Akimasa Kataoka}
\affiliation{Division of Science, National Astronomical Observatory of Japan,
2-21-1 Osawa, Mitaka, Tokyo 181-8588, Japan}

\begin{abstract} 

{\fourthcomment{We introduce a possible disruption mechanism of dust grains in planet formation by their spinning motion.}}
This mechanism has been discussed as rotational disruption for the interstellar dust grains.
We theoretically calculate whether porous dust aggregates can be disrupted by their spinning motion and if it prohibits dust growth in protoplanetary disks.
We assume radiative torque and gas-flow torque as driving sources of the spinning motion, assume that dust aggregates reach a steady-state rigid rotation, and compare the obtained tensile stress due to the centrifugal force with their tensile strength.
{\fourthcomment{We model the irregularly-shaped dust aggregates by introducing a parameter, $\gamma_\mathrm{ft}$, that mimics the conversion efficiency from force to torque.}}
As a result, we find that porous dust aggregates are rotationally disrupted by their spinning motion induced by gas flow when their mass is larger than $\sim10^8$ g and their volume filling factor is smaller than $\sim 0.01$ in our fiducial model, while relatively compact dust aggregates with volume filling factor more than 0.01 do not face this problem.
If we assume the dust porosity evolution, we find that dust aggregates whose Stokes number is $\sim0.1$ can be rotationally disrupted in their growth and compression process.
Our results suggest that the growth of dust aggregates may be halted due to rotational disruption or that other compression mechanisms are needed to avoid it.
{\fifthcomment{We also note that dust aggregates are not rotationally disrupted when $\gamma_\mathrm{ft}\leq0.02$ in our fiducial model and the modeling of irregularly-shaped dust aggregates is essential in future work.}}

\end{abstract}


\section{Introduction} \label{sec:intro}

Planets form from submicrometer-sized dust grains in protoplanetary disks.
Kilometer-sized planetesimals are thought to form during the planet formation.
Submicrometer-sized dust grains called monomers collide and grow in the direct coagulation process.
This process naturally makes porous dust aggregates, which have been investigated theoretically and experimentally \citep[e.g.,][]{Ossenkopf1993,Dominik1997,Blum2000,Wada2007,Wada2008,Wada2009,Suyama2008,Okuzumi2012,Kataoka2013L}. 
If a dust layer becomes unstable during the dust growth, dust grains are concentrated rapidly and form larger bodies, which are called planetesimals.
There are several instabilities such as gravitational instability \citep[e.g.,][]{Goldreich1973} and streaming instability \citep[e.g.,][]{Youdin2005,Johansen2007,Johansen2011} to form planetesimals.

Dust grains can be collisionally disrupted during their growth, which can explain observed protoplanetary disks.
The small spectral index of protoplanetary disks infers that the maximum size of dust grains is from millimeter to centimeter \citep[e.g.,][for a review]{Testi2014}.
Also, some polarization observations show that the maximum grain size is $\sim100\mathrm{\ \mu m}$ \citep[e.g.,][]{Kataoka2016HLTau,Kataoka2016HD,Kataoka2017,Stephens2017}.
This maximum grain size can be obtained if the polarization mechanism is self-scattering \citep{Kataoka2015}.
The self-scattering can be observed at the wavelength which is almost the same as the maximum grain size.
Fragmentation is required to make such submillimeter-sized dust grains \citep[e.g.,][]{Birnstiel2011}.
Catastrophic collision between two dust grains leads to such fragmentation \citep[e.g.,][]{Blum1993}.
In this paper, we discuss another possible barrier of dust growth in planet formation, which is called rotational disruption.

Rotational disruption of dust grains has been investigated in context of the interstellar space \citep[e.g.,][]{Hoang2019ApJ,Hoang2019NatAs}.
In the interstellar space, dust grains receive a strong radiation field from a young massive star-cluster or a supernova.
The radiation field causes radiative torque of dust grains, which leads to their spinning motion.
Spinning dust grains feel the centrifugal force, which is the tensile stress.
When the tensile stress is larger than the tensile strength of dust grains, they are disrupted.

\begin{figure*}
\plotone{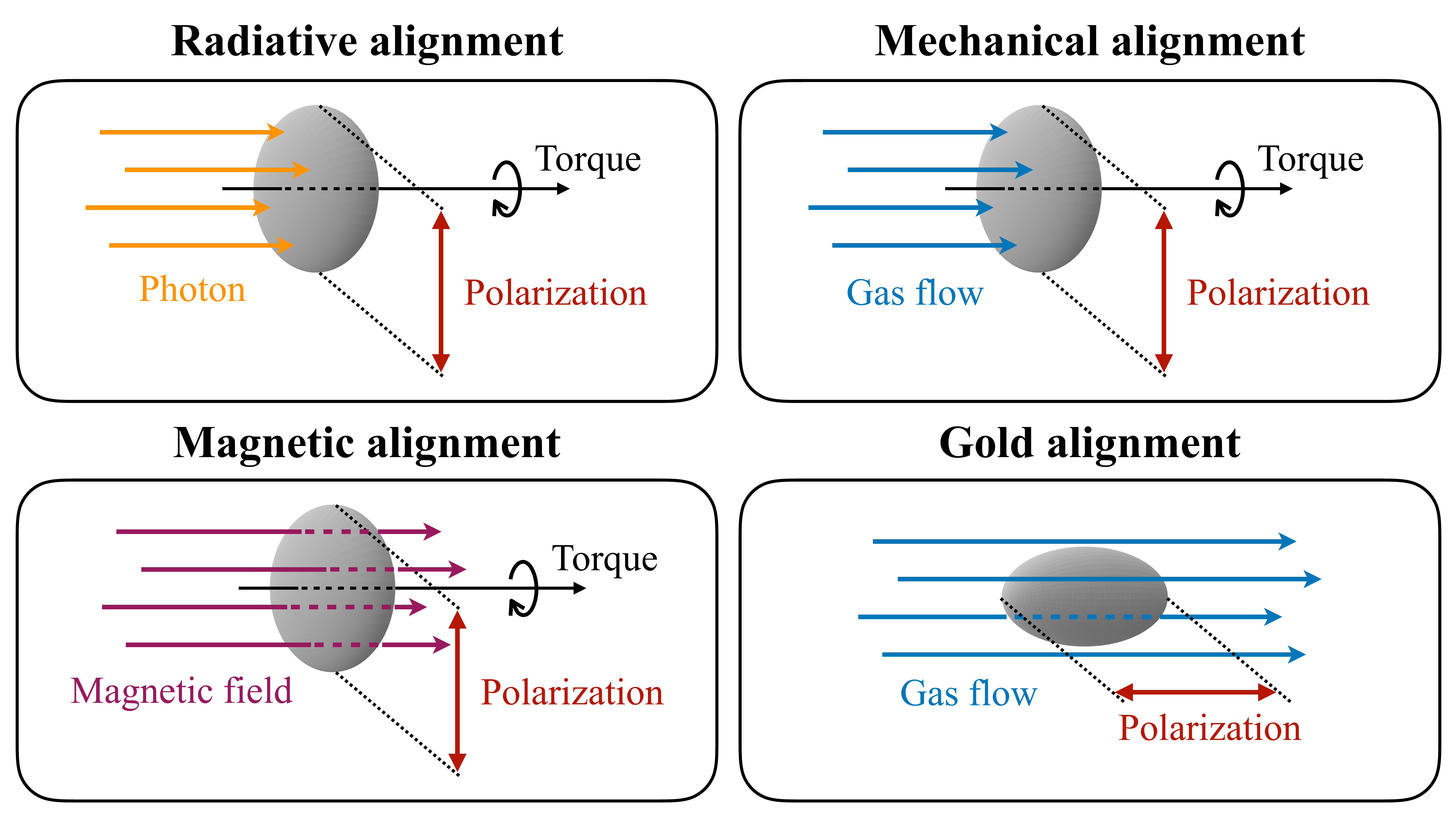}
\caption{Dust-alignment mechanisms in protoplanetary disks: radiative, mechanical, magnetic, and Gold alignment.
The radiative, mechanical, and magnetic alignment require some torque, while Gold alignment does not require any torque.
The spin axis, which is parallel to the minor axis of an elliptical dust grain, tends to be parallel to the flow of photon or gas, or the direction of the magnetic field.
The elliptical dust grain emits thermal polarization waves, which is parallel to its major axis.
In this paper, we assume alignments which require some rotational torque: radiative torque from photon's spin angular momentum and momentum, and gas-flow torque, which is also called the mechanical torque in the previous works.
\label{fig:alignment}}
\end{figure*}

Spinning motion of dust grains is inferred by observations of protoplanetary disks.
The spinning dust grains tend to align and emit thermal polarized waves {\mtatsuuma{\citep[e.g.,][]{Draine1996,Draine1997,Lazarian2007RAT,Lazarian2007MA,Hoang2008,Hoang2009Pinwheel,Hoang2009RAT,Hoang2014,Lazarian2015,Tazaki2017,Hoang2018,Kataoka2019}}}.
The polarized emission due to such dust alignment in protoplanetary disks has been observed at radio wavelength \citep[e.g.,][]{Kataoka2017,Stephens2017}.
The direction of polarization vectors is under debate \citep[e.g.,][]{Yang2019} because the alignment mechanism in protoplanetary disks is unclear.
The possible alignment mechanism is as follows \citep[Figure \ref{fig:alignment}; for details see Section 2 in][]{Kataoka2019}.
When irregularly-shaped dust grains receive anisotropic radiation field or some gas flow, they begin to spin-up.
We call these spin-up torques radiative and gas-flow torque.
The gas-flow torque is the same as the mechanical torque in the previous works.
Then, the spin axis becomes parallel to the radiation, gas flow, or magnetic-field direction, which we call radiative, mechanical, or magnetic alignment, respectively.
Although Gold alignment \citep{Gold1952} does not require any rotation, we assume that dust grains align due to some spinning motion in this paper because the relative velocity between gas and dust is subsonic in protoplanetary disks.
{\mtatsuuma{Aside from the alignment mechanism in protoplanetary disks, it is inferred that the spinning motion of dust grains is required to explain some radio polarization observations, as mentioned above in this paragraph.}}

In this paper, we apply the concept of the rotational disruption \citep[e.g.,][]{Hoang2019ApJ,Hoang2019NatAs} to the case of porous dust aggregates during their growth in protoplanetary disks.
First, we explain how to investigate whether dust aggregates can be rotationally disrupted in Section \ref{sec:methods}.
We assume radiative torque \citep[e.g.,][]{Draine1996} and gas-flow torque.
{\misako{The situation in protoplanetary disks would be completely different from that in the interstellar space because of the grain growth and the dense gas.}}
{\mtatsuuma{During dust aggregate growth, their volume filling factor decreases to $\sim10^{-5}$ and their size increases to km-scale \citep{Kataoka2013L}.}}
{\misako{Therefore, we treat both the radiative torque and the gas-flow torque.}}
{\mtatsuuma{The gas-flow torque depends on the helicity of dust aggregates, which has not been investigated in the context of the grain growth in the planet formation, although some previous studies show that irregular grains may not show well-defined helicity \citep[e.g.,][]{Lazarian2015}.
As the first step and for simplicity, we introduce a simple parameter, force-to-torque efficiency, to investigate the effect of the helicity of dust aggregates in Section \ref{subsubsec:gastorque}.}}
{\misako{We also assume that dust aggregates reach a steady-state rigid rotation.}}
By comparing the obtained tensile stress due to the centrifugal force and tensile strength of dust aggregates \citep{Tatsuuma2019}, we find out whether they can be rotationally disrupted.
Next, we show our results in Section \ref{sec:results}, which contains the mass and volume filling factor of rotationally disrupted dust aggregates.
We also investigate the dependence of our results on physical parameters.
Then, we interpret and discuss our results in Section \ref{sec:discussion}.
Finally, we conclude our work in Section \ref{sec:conclusions}.

\section{Methods} \label{sec:methods}

In this section, we explain how we calculate whether dust aggregates are rotationally disrupted during their growth in protoplanetary disks.
We treat a dust aggregate as a rigid sphere.
We assume that dust aggregates receive spin-up torques and spin-down torques, which leads to the steady-state angular velocity of the rotation.
We assume two origins of spin-up torques: radiation \citep[e.g.,][]{Draine1996} and gas flow, which we call radiative and gas-flow torque and explain them in Section \ref{subsec:spin-up}.
The gas-flow torque is known as the mechanical torque.
Also, we explain the spin-down torque due to surrounding gas in Section \ref{subsec:spin-down}.
When the spin-up torques and the spin-down torque are balanced, the spinning motion of dust aggregates becomes steady-state.
The steady-state angular velocity is described in Section \ref{subsec:angvel}.
Then, we calculate the tensile stress due to the centrifugal force \citep[e.g.,][]{Hoang2019ApJ,Hoang2019NatAs}, which we explain in Section \ref{subsec:stress}.

Models to calculate the tensile stress and tensile strength of dust aggregates in protoplanetary disks are as follows.
To calculate the spin-up torques, we need models of the radiation field and gas disk, which are explained in Section \ref{subsec:diskmodel}.
The radiation field at the midplane is dominated by dust thermal emission, which was investigated by \citet{Tazaki2017}.
We adopt their model and assume the minimum mass solar nebula.
The tensile strength of dust aggregates was investigated by \citet{Tatsuuma2019} using three-dimensional dust N-body simulations, which we explain in Section \ref{subsec:strength}.
To investigate whether the rotational disruption affects dust growth, we explain the model of porosity evolution in Section \ref{subsec:dustgrowth}.
The porosity evolution of dust aggregates was investigated by \citet{Kataoka2013L}, who assume hit-and-stick growth, collisional compression, gas compression, and self-gravitational compression.

\subsection{Spin-up Torques} \label{subsec:spin-up}

We assume two origins of spin-up torques in protoplanetary disks: radiation and gas flow, which we call radiative and gas-flow torque (see Figure \ref{fig:alignment}).
The gas-flow torque is also known as the mechanical torque.
The radiative torque originates from both spin angular momentum and momentum of photons, which we describe in Section \ref{subsubsec:radtorque}.
Photons have spin angular momentum, which is described as $\hbar$.
This spin angular momentum may be directly converted to the radiative torque.
Also, photons have momentum, which is described as $h\nu /c$.
If dust aggregates are not spherical but irregularly shaped, they may receive torque by the photon momentum.
The irregularly-shaped dust aggregates may also receive torque by gas flow, which we describe using gas drag force in Section \ref{subsubsec:gastorque}.

{\fourthcomment{Another possible source of the spin-up torques would be caused by an anisotropic distribution of temperature of an object, which is generally known as ``the Crookes light mill'' effect.}}
{\tatsuuma{When one side of the object is heated by infrared radiation, surrounding gas molecules give larger momentum to the heated side than to the non-heated side.
This difference of momentum may cause spin-up torque, which is different from the radiative torques and the gas-flow torque.
In this work, we do not treat the temperature distribution of a dust aggregate for simplicity, so that we do not include this effect.}}

\subsubsection{Radiative Torques} \label{subsubsec:radtorque}

Radiative torque due to spin angular momentum of photons is defined as \citep{Draine1996} 
\begin{equation}
\Gamma_\mathrm{up,rad,s} = \pi a^2\gamma_\mathrm{rad} u_\mathrm{rad}\frac{\lambda}{2\pi}Q_\Gamma,
\label{eq:Gammauprads}
\end{equation}
where $a$ is the radius of dust aggregates, $\gamma_\mathrm{rad}$ is the anisotropy parameter of the radiation field, $u_\mathrm{rad}$ is the energy density of the radiation field, $\lambda$ is the radiation wavelength, and $Q_\Gamma$ is the radiative-torque efficiency.
The concept of Equation (\ref{eq:Gammauprads}) is as follows.
The number of photons hitting a dust aggregate per unit time per unit area is $cu_\mathrm{rad}\gamma_\mathrm{rad}/(h\nu)$ because the net Poynting flux is $cu_\mathrm{rad}\gamma_\mathrm{rad}$ and the energy of a single photon is $h\nu$.
Each photon carries the spin angular momentum $\hbar=h/(2\pi)$ and we assume that the efficiency for converting it into torque is $Q_\Gamma$.
Because the cross-section of the dust aggregate is $\pi a^2$, the radiative torque due to spin angular momentum of photons is described as $\pi a^2\cdot cu_\mathrm{rad}\gamma_\mathrm{rad}/(h\nu)\cdot h/(2\pi)\cdot Q_\Gamma$.

\begin{figure}
\plotone{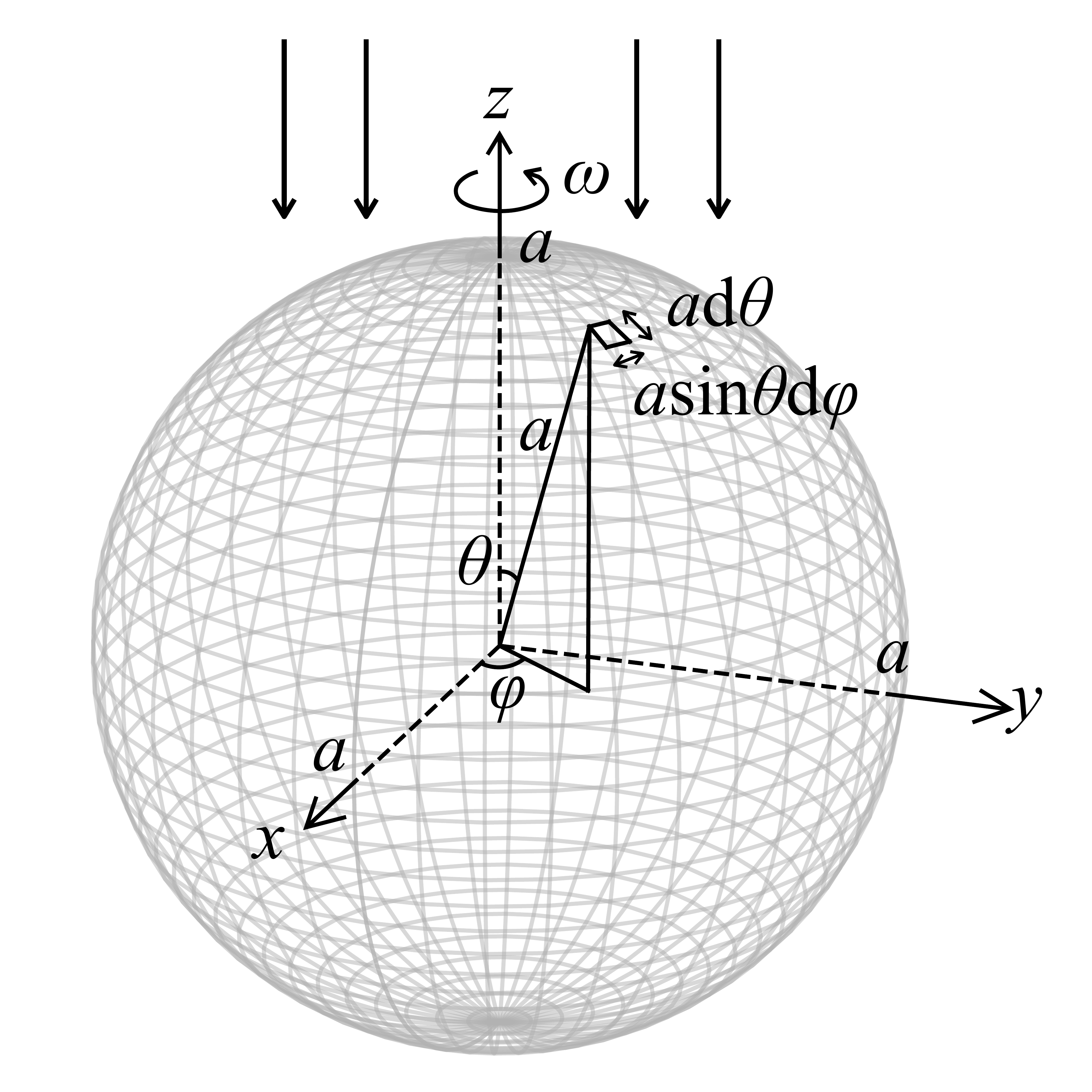}
\caption{Upper hemisphere of a dust aggregate which receives the photon momenta or the gas drag force.
The spin axis is parallel to the direction of the photon momenta or the gas flow.
The unit area of the hemisphere is described as $a^2\sin\theta\mathrm{d}\theta\mathrm{d}\varphi$. \label{fig:hemisphere}}
\end{figure}

In addition, we introduce radiative torque due to photon momentum as
\begin{equation}
\Gamma_\mathrm{up,rad,p} = \frac{2}{3}\pi a^3\gamma_\mathrm{p}u_\mathrm{rad}Q_\Gamma,
\label{eq:Gammaupradp}
\end{equation}
where $\gamma_\mathrm{p}$ is the anisotropy parameter related to receiving photon momentum.
We note that the radiative torque due to photon momentum dominates over that due to spin angular momentum in protoplanetary disks because the radius of dust aggregates is much larger than that of dust grains in the interstellar medium.
The concept of Equation (\ref{eq:Gammaupradp}) is as follows.
Consider the upper hemisphere of a dust aggregate (see Figure \ref{fig:hemisphere}) which receives the photon momenta and assume $\gamma_\mathrm{p}$ times all momenta are converted into torque.
The total number of photons hitting the dust aggregate per unit time per unit area is $cu_\mathrm{rad}\gamma_\mathrm{p}/(h\nu)$ and the photon momentum is $h\nu/c$.
The radiative torque due to photon momentum is described as $\int_0^{2\pi}\int_0^{\pi/2}cu_\mathrm{rad}\gamma_\mathrm{p}/(h\nu)\cdot h\nu/c\cdot Q_\Gamma\cdot\cos\theta\cdot a\sin\theta \cdot a^2\sin\theta\mathrm{d}\theta\mathrm{d}\varphi$.

In this work, we introduce the expression of the radiative-torque efficiency $Q_\Gamma$ for porous dust aggregates as
\begin{equation}
Q_\Gamma \sim \left\{
\begin{aligned}
& 0.4\left(\frac{1.8a\phi}{\lambda}\right)^3 & (\lambda>1.8a\phi)\\
& 0.4 & (\lambda\le1.8a\phi)
\end{aligned} \right.,
\label{eq:Qrad}
\end{equation}
where $\phi$ is the volume filling factor of dust aggregates \citep[e.g.,][]{Lazarian2007RAT}.
This radiative-torque efficiency can be divided into two factors: absorption coefficient $Q_\mathrm{abs}$ and torque efficiency, that is $Q_\Gamma=Q_\mathrm{abs}\times\mathrm{(torque\ efficiency)}$.
The absorption coefficient $Q_\mathrm{abs}$ of porous dust aggregates can be characterized by $a\phi$ \citep{Kataoka2014}.
The power-law index and the constant value are fitted by \citet{Lazarian2007RAT}.
They derived $Q_\Gamma$ as a function of $\lambda/a$ by using numerical simulations.
We adopt the formulation that $Q_\Gamma\sim0.4(1.8a/\lambda)^3$ for $\lambda>1.8a$ and $Q_\Gamma\sim0.4$ for $\lambda\le1.8a$ and newly expand it into the case of porous dust aggregates by replacing $a$ with $a\phi$.
It is important that the radiative-torque efficiency $Q_\Gamma$ become smaller for longer wavelength and the boundary wavelength is defined as $\lambda\sim a\phi$.

\subsubsection{Gas-flow torque} \label{subsubsec:gastorque}

We also introduce gas-flow torque as
\begin{equation}
\Gamma_\mathrm{up,gas} = \frac{2}{3}aF_\mathrm{drag}\gamma_\mathrm{ft},
\label{eq:Gammaupgas}
\end{equation}
where $F_\mathrm{drag}$ is the gas drag force received by a dust aggregate and $\gamma_\mathrm{ft}$ is the force-to-torque efficiency, which we introduce in this work and is related to the helicity of dust aggregates.
{\mtatsuuma{Although we treat a dust aggregate as a rigid sphere at first, we introduce this efficiency to include the effect of non-spherical grains.}}
The concept of Equation (\ref{eq:Gammaupgas}) is as follows.
Consider the upper hemisphere of a dust aggregate (see Figure \ref{fig:hemisphere}) which receives the gas drag force.
The drag force per unit area is assumed to be $F_\mathrm{drag}/(\pi a^2)$.
If all of the drag force is converted into torque, the gas-flow torque is described as $\int_0^{2\pi}\int_0^{\pi/2}F_\mathrm{drag}/(\pi a^2)\cdot\cos\theta\cdot a\sin\theta \cdot a^2\sin\theta\mathrm{d}\theta\mathrm{d}\varphi=(2/3)aF_\mathrm{drag}$.
In reality, not all of the drag force is converted into torque.
Thus, we introduce the force-to-torque efficiency $\gamma_\mathrm{ft}$.

{\tatsuuma{We assume that the direction of the gas-flow torque is always the same for simplicity, similar to leaves rotating and falling from a tree.}}
{\fourthcomment{However, this demonstrative phenomenon is happening on Earth's air, which has high viscosity. Since the viscosity in disks is low, it is unclear for dust aggregates to be able to use the same analogy.}}

We explain how to calculate the drag force, which is defined as
\begin{equation}
F_\mathrm{drag}=\frac{mv}{t_\mathrm{s}},
\label{eq:Fdrag}
\end{equation}
where $m$ is the mass of dust aggregates, $v$ is the relative velocity between dust aggregates and gas, and $t_\mathrm{s}$ is the stopping time.
The stopping time is defined as 
\begin{equation}
t_\mathrm{s}=\left\{
\begin{aligned}
& \frac{\rho a}{\rho_\mathrm{gas}v_\mathrm{th}} & \left(a<\frac{9\lambda_\mathrm{mfp}}{4}\right)\\
& \frac{8\rho a}{3C_\mathrm{D}\rho_\mathrm{gas}v} & \left(a>\frac{9\lambda_\mathrm{mfp}}{4}\right)
\end{aligned} \right.,
\label{eq:stoppingtime}
\end{equation}
where $\rho$ is the mean internal density of dust aggregates, $\rho_\mathrm{gas}$ is the gas density, $v_\mathrm{th}=\sqrt{8/\pi}c_\mathrm{s}$ is the thermal velocity, $c_\mathrm{s}$ is the sound velocity, $C_\mathrm{D}$ is the dimensionless drag coefficient, and $\lambda_\mathrm{mfp}$ is the mean free path of gas molecules.
The former of Equation (\ref{eq:stoppingtime}) is known as the Epstein regime, while the latter is known as the Stokes and Newton regimes.
The dimensionless drag coefficient is described as \citep[e.g.,][]{Weidenschilling1977}
\begin{equation}
C_\mathrm{D}\simeq\left\{
\begin{aligned}
& 24\Rey^{-1} & (\Rey<1)\\
& 24\Rey^{-0.6} & (1<\Rey<800)\\
& 0.44 & (\Rey>800)
\end{aligned} \right.,
\label{eq:CD}
\end{equation}
where $\Rey=2av/\nu_\mathrm{mol}$ is the Reynolds number and $\nu_\mathrm{mol}=v_\mathrm{th}\lambda_\mathrm{mfp}/2$ is the molecular viscosity.

We assume that the relative velocity between dust aggregates and gas is divided into radial and azimuthal components as
\begin{eqnarray}
v_r &=& \frac{2\St}{1+\St^2}\eta v_\mathrm{K} \label{eq:vradial}\\
v_\varphi &=& \frac{\St^2}{1+\St^2}\eta v_\mathrm{K}, \label{eq:vazimuth}
\end{eqnarray}
where $\St=\Omega_\mathrm{K}t_\mathrm{s}$ is the Stokes number, $\Omega_\mathrm{K}$ is the Keplerian angular velocity, and $\eta v_\mathrm{K}$ is the gas rotational velocity relative to the Keplerian velocity $v_\mathrm{K}$ \citep[e.g.,][]{Adachi1976,Nakagawa1986}.
We assume that there is no turbulence for simplicity, and therefore the relative velocity is given as
\begin{equation}
v=\sqrt{v_r^2+v_\varphi^2}.
\label{eq:v}
\end{equation}

\subsection{Spin-down Torque} \label{subsec:spin-down}

We assume that the spin-down torque originates from elastic collisions of surrounding gas molecules.
Other spin-down torques can be negligible because the gas density in protoplanetary disks is much larger than that of the interstellar medium.

\begin{figure}
\plotone{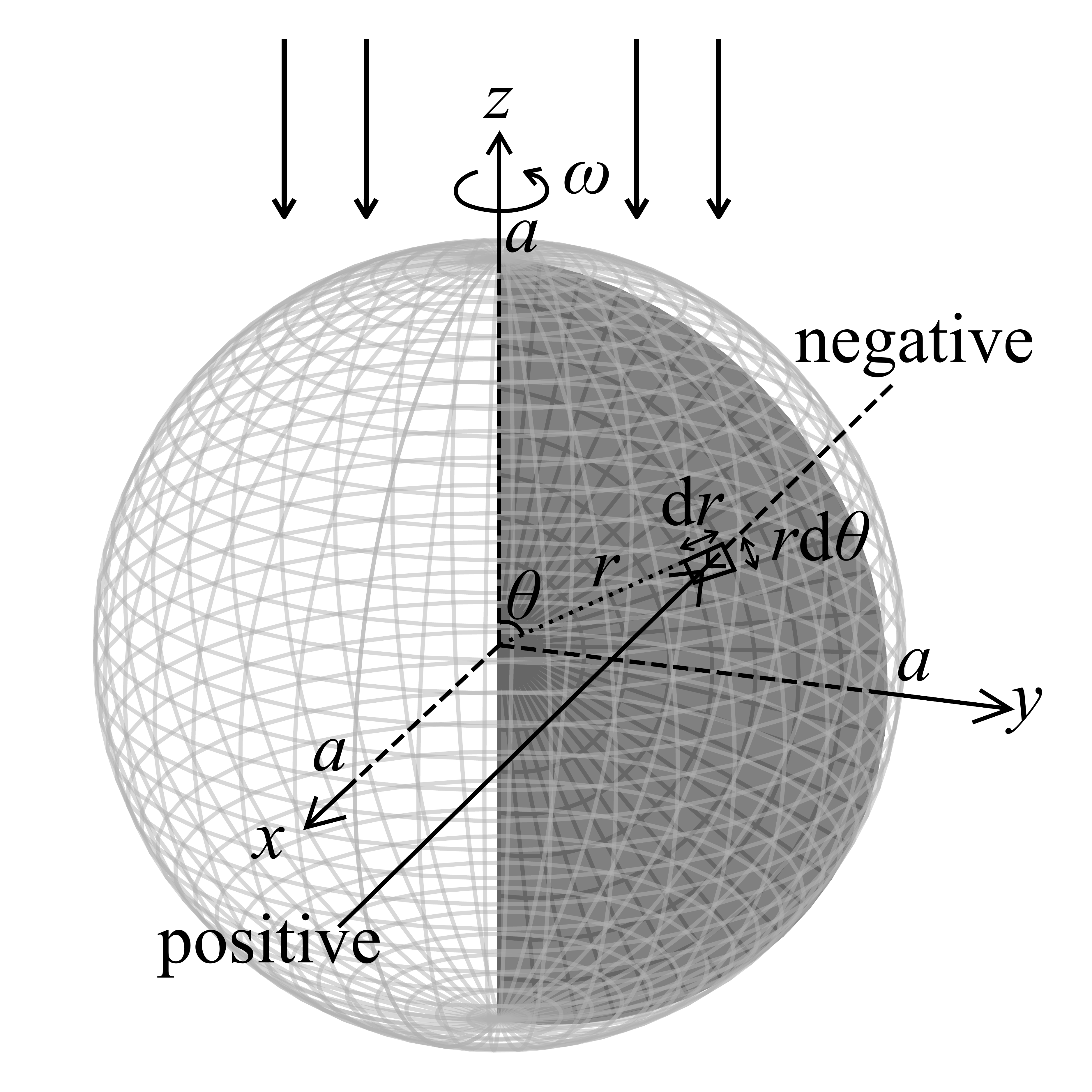}
\caption{An arbitrary semicircle of a cross-section of a dust aggregate.
The gas molecule going backward gives positive torque to the dust aggregate, while another gas molecule going forward gives negative torque to it.
The unit area of the semicircle is described as $r\mathrm{d}\theta\mathrm{d}r$.
\label{fig:semicircle}}
\end{figure}

The spin-down torque can be written as
\begin{equation}
\Gamma_\mathrm{down}\sim-\frac{I\omega}{t_\mathrm{s}},
\label{eq:Gammadown}
\end{equation}
where $I=8\pi\rho a^5/15$ is the inertia moment of a dust aggregate and $\omega$ is the angular velocity.
This means that the angular momentum of the dust aggregate $I\omega$ is transferred to the surrounding gas in a timescale $t_\mathrm{s}$.
The concept of Equation (\ref{eq:Gammadown}) is the same as the Epstein drag.
Consider a spherical dust aggregate rotating at an angular velocity $\omega$ and a semicircle of a cross-section of the dust aggregate (see Figure \ref{fig:semicircle}).
A gas molecule collides with it elastically and gives it a negative torque or a positive torque dependent on the direction where the gas molecule comes from.
The total negative torque is $-2\int_0^a\int_{-\pi/2}^{\pi/2}\rho_\mathrm{gas}/m_\mathrm{gas}\cdot(\omega r\cos\theta+v_\mathrm{th})\cdot2m_\mathrm{gas}v_\mathrm{th}\cdot r\cos\theta\cdot r\mathrm{d}\theta\mathrm{d}r$, while the total positive torque is $2\int_0^a\int_{-\pi/2}^{\pi/2}\rho_\mathrm{gas}/m_\mathrm{gas}\cdot(-\omega r\cos\theta+v_\mathrm{th})\cdot2m_\mathrm{gas}v_\mathrm{th}\cdot r\cos\theta\cdot r\mathrm{d}\theta\mathrm{d}r$, where $m_\mathrm{gas}$ is the molecular mass.
Therefore, the net torque due to elastic collision of surrounding gas molecules is $\Gamma_\mathrm{down}=-\rho_\mathrm{gas}v_\mathrm{th}a^4\omega$.
In the Epstein regime, in other words, if $a<9\lambda_\mathrm{mfp}/4$, the spin-down torque can be written with the inertia moment $I$ and the stopping time $t_\mathrm{s}$ as $\Gamma_\mathrm{down}=-15I\omega/(8\pi t_\mathrm{s})\sim-I\omega/t_\mathrm{s}$.
We take a rough approximation to apply the spin-down torque to the Stokes and Newton regimes.

{\tatsuuma{Although we use the stopping time $t_\mathrm{s}$ to apply the spin-down torque (Equation (\ref{eq:Gammadown})) to the Stokes and Newton regimes, it is unclear that this approximation is accurate.
In these regimes, we treat gas molecules as a fluid, not as particles, so that we have to calculate the fluid dynamics.
In this work, however, we assume the spin-down torque in the Epstein regime for simplicity.}}

\subsection{Steady-state Angular Velocity} \label{subsec:angvel}

In the steady-state, the total torque is zero, in other words,
\begin{equation}
\frac{\mathrm{d}\omega}{\mathrm{d}t}=\frac{\Gamma_\mathrm{up}+\Gamma_\mathrm{down}}{I}=0,
\label{eq:steadystate}
\end{equation}
where $\Gamma_\mathrm{up}=\Gamma_\mathrm{up,rad,s}+\Gamma_\mathrm{up,rad,p}+\Gamma_\mathrm{up,gas}$ is the total spin-up torque.
Equation (\ref{eq:Gammadown}) and Equation (\ref{eq:steadystate}) show that $\Gamma_\mathrm{up}=-\Gamma_\mathrm{down}=I\omega_\mathrm{c}/t_\mathrm{s}$, where $\omega_\mathrm{c}$ is the steady-state angular velocity, and it can be written as
\begin{equation}
\omega_\mathrm{c} = \frac{\Gamma_\mathrm{up}t_\mathrm{s}}{I} = \frac{(\Gamma_\mathrm{up,rad,s}+\Gamma_\mathrm{up,rad,p}+\Gamma_\mathrm{up,gas})t_\mathrm{s}}{I}.
\label{eq:constangvel}
\end{equation}

\subsection{Tensile Stress due to Centrifugal Force} \label{subsec:stress}

\begin{figure}
\plotone{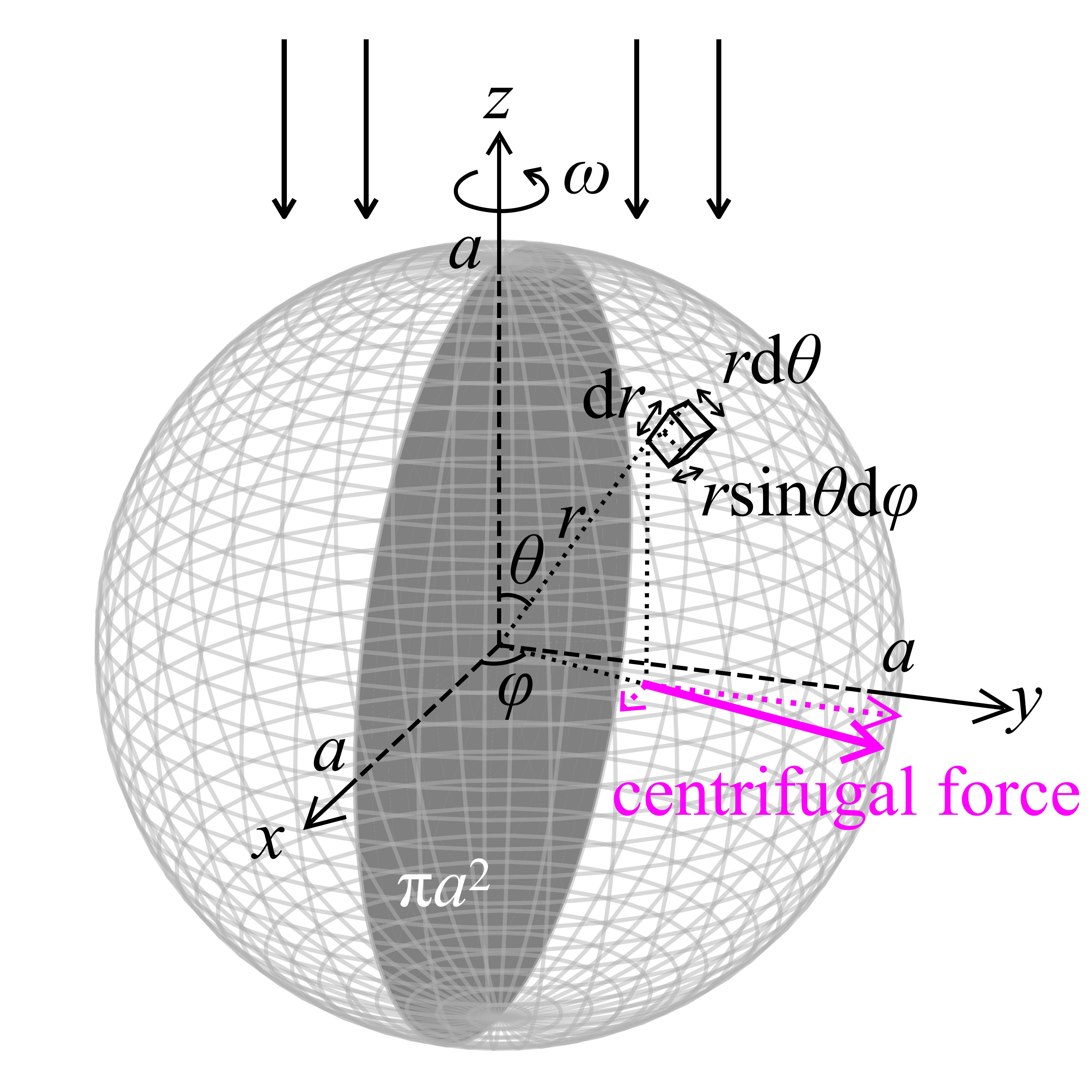}
\caption{An arbitrary hemisphere of a dust aggregate whose cut section includes the spin axis ($z$-axis).
The unit volume of the hemisphere is described as $r^2\sin\theta\mathrm{d}r\mathrm{d}\theta\mathrm{d}\varphi$.
The solid magenta arrow represents the centrifugal force felt by the unit volume, which is divided into $x$ and $y$ components (dotted magenta arrows).
To derive the tensile stress, we focus on the $y$ component of the centrifugal force.
\label{fig:hemispherevolume}}
\end{figure}

Average tensile stress due to centrifugal force of a rotating sphere with the constant angular velocity $\omega_\mathrm{c}$ is \citep[e.g.,][]{Hoang2019NatAs}
\begin{equation}
S = \frac{\rho a^2\omega_\mathrm{c}^2}{4}.
\label{eq:stress}
\end{equation}
The concept of Equation (\ref{eq:stress}) is as follows.
Consider the total $y$-component centrifugal force of a hemisphere (see Figure \ref{fig:hemispherevolume}).
This is defined as $\int_0^a\int_0^\pi\int_{0}^{\pi}\rho\omega_\mathrm{c}\cdot r\sin\theta\omega_\mathrm{c}\cdot\sin\varphi\cdot r^2\sin\theta\mathrm{d}\varphi\mathrm{d}\theta\mathrm{d}r=\pi\rho a^4\omega_\mathrm{c}^2/4$.
This force is applied to the area $\pi a^2$, and thus the tensile stress is $\rho a^2\omega_\mathrm{c}^2/4$.

\subsection{Disk Model} \label{subsec:diskmodel}

In this section, we explain models of the radiation field and gas disk to calculate the spin-up and spin-down torques.
We focus on the midplane of protoplanetary disks, where dust aggregates settle and planetesimals form.
The radiation field there is dominated by dust thermal emission.
We adopt the values related to radiative torques according to \citet{Tazaki2017} as
\begin{eqnarray}
\lambda &\simeq& 140\mathrm{\ \mu m}\label{eq:lambda}\\
u_\mathrm{rad} &\simeq& 5.2\times10^{-8}(R/10\mathrm{\ au})^{-2}\mathrm{\ erg\ cm^{-3}}\label{eq:urad}\\
\gamma_\mathrm{rad} &\simeq& \left\{
\begin{aligned}
& 0.1 & (R\lesssim30\mathrm{\ au})\\
& 1 & (R\gtrsim50\mathrm{\ au})
\end{aligned} \right.,\label{eq:anisotropy}
\end{eqnarray}
where $R$ is the orbital radius.
They investigated the radiative alignment due to the radiative torque by using the three-dimensional radiative transfer calculation.
The radiation wavelength $\lambda\simeq140\mathrm{\ \mu m}$ means that thermal emission from cold dust aggregates ($\sim20$ K) is dominant at the midplane.
We assume there is only one wavelength in the radiation field for simplicity.
We assume that the energy density of the radiation field $u_\mathrm{rad}$ is $10^{-2}$ times lower than that at the surface layer of the disk.
The anisotropy parameter of the radiation field $\gamma_\mathrm{rad}$ shows that the disk is optically thick at $R\lesssim30\mathrm{\ au}$, while optically thin at $R\gtrsim50\mathrm{\ au}$.

We use the minimum-mass solar nebula (MMSN) model \citep{Hayashi1981} with a solar-mass central star for the model of the gas disk as follows.
The gas surface density is $\Sigma_\mathrm{gas}=1700(R/1\mathrm{\ au})^{-3/2}\mathrm{\ g\ cm^{-2}}$.
The disk temperature is $T=280(R/1\mathrm{\ au})^{-1/2}\mathrm{\ K}$.
The gas density is $\rho_\mathrm{gas}=\Sigma_\mathrm{gas}/(\sqrt{2\pi}h_\mathrm{gas})$, where $h_\mathrm{gas}=c_\mathrm{s}/\Omega_\mathrm{K}$ is the gas scale height.
The sound velocity is $c_\mathrm{s}=\sqrt{k_\mathrm{B}T/m_\mathrm{gas}}$, where $k_\mathrm{B}$ is the Boltzmann constant and $m_\mathrm{gas}=3.9\times10^{-24}\mathrm{\ g}$ is the mean molecular mass.
The Keplerian angular velocity is $\Omega_\mathrm{K}=\sqrt{GM_\odot/R^3}$, where $G$ is the gravitational constant and $M_\odot$ is the solar mass.
The mean free path of gas molecules is $m_\mathrm{gas}/(\sigma_\mathrm{gas}\rho_\mathrm{gas})$, where $\sigma_\mathrm{gas}=2\times10^{-15}\mathrm{\ cm^2}$ is the collision cross-section of gas molecules.
In this model, $\eta v_\mathrm{K}=54\mathrm{\ m\ s^{-1}}$ is the constant through all orbital radii.

\subsection{Tensile Strength of Porous Dust Aggregates} \label{subsec:strength}

We need to compare the tensile stress due to the centrifugal force and their tensile strength, which we explain in this section, to find out whether dust aggregates can be rotationally disrupted.
The tensile strength of porous dust aggregates was investigated by using three-dimensional dust-N-body simulations \citep{Tatsuuma2019} and is given as
\begin{equation}
S_\mathrm{max} \sim 6\times10^5\left(\frac{\gamma_\mathrm{s}}{100\mathrm{\ mJ\ m^{-2}}}\right)\left(\frac{r_0}{0.1\mathrm{\ \mu m}}\right)^{-1}\phi^{1.8}\mathrm{\ Pa},
\label{eq:tensilestrength}
\end{equation}
where $\gamma_\mathrm{s}$ is the surface energy of monomers and $r_0$ is the monomer radius.
They calculated interactions of connections between two monomers in contact.
In this paper, we assume ice dust aggregates, which means $\gamma_\mathrm{s}=100\mathrm{\ mJ\ m^{-2}}$.

The interpretation of the tensile strength of porous dust aggregates is as follows.
Equation (\ref{eq:tensilestrength}) can be derived from the maximum force needed to separate two sticking-monomers and the sub-structure radius of dust aggregates.
This is different from the tensile strength of \citet{Greenberg1995}, who derived it by assuming that all connections between monomers are broken.
In protoplanetary disks, dust aggregates are much larger than dust grains in the interstellar medium, and thus this assumption might not be suitable for this situation.

\subsection{Porosity Evolution of Dust Aggregates} \label{subsec:dustgrowth}

In this section, we explain the model of porosity evolution of dust aggregates to investigate whether the rotational disruption affects the dust growth.
We calculate the volume filling factor for a given mass of dust aggregates during their growth.
In other words, we take the maximum volume filling factor among those determined by their fractal dimension, ram pressure of disk gas, and self-gravitational pressure when their compressive strength equals the given pressure \citep{Kataoka2013L}.
We neglect collisional compression, Brownian motion of dust aggregates, and turbulence of disk gas for simplicity.
We assume that the collisional compression and the Brownian motion do not affect the maximum volume filling factor.
If there is no turbulence, however, we might underestimate the maximum volume filling factor determined by the ram pressure.
We discuss the strength of turbulence in Section \ref{subsec:turbulence}.

Initially, dust aggregates grow via hit-and-stick collisions, where their fractal dimension is $\sim2$ \citep[e.g.,][]{Okuzumi2012}.
The number of monomers $N$ of a dust aggregate with a fractal dimension $D$ satisfies $N\propto a^D$ and $\phi=Nr_0^3/a^3$, and therefore we obtain $\phi\propto m^{(D-3)/D}$.
The volume filling factor of dust aggregates with a fractal dimension of 2 is given as
\begin{equation}
\phi_{D=2} = \left(\frac{m}{m_0}\right)^{-1/2},
\label{eq:hitandstickphi}
\end{equation}
where $m_0=(4/3)\pi r_0^3\rho_0$ is the monomer mass and $\rho_0$ is the material density, because the initial condition is $\phi=1$ when $m=m_0$.
The material density of ice is $\rho_0=1.0\mathrm{\ g\ cm^{-3}}$.

As dust aggregates grow, they begin to be compressed by disk gas and their self-gravity.
The compressive strength is given as \citep{Kataoka2013}
\begin{equation}
P_\mathrm{comp} = \frac{E_\mathrm{roll}}{r_0^3}\phi^3,
\label{eq:compstrength}
\end{equation}
where $E_\mathrm{roll}=6\pi^2\gamma_\mathrm{s}r_0\xi_\mathrm{crit}$ is the rolling energy needed to rotate a monomer around its connection point by $90^\circ$ and $\xi_\mathrm{crit}$ is the critical rolling displacement \citep[e.g.,][]{Dominik1997,Wada2007}.
The theoretical critical rolling displacement ($\xi_\mathrm{crit}=2\textrm{\ \AA}$, \citealt{Dominik1997}) is different from the experimental one ($\xi_\mathrm{crit}=32\textrm{\ \AA}$, \citealt{Heim1999}).
We adopt $\xi_\mathrm{crit}=8\textrm{\ \AA}$ in this paper{\mtatsuuma{, whose value is the same as \citet{Kataoka2013L}.}}

The ram pressure of disk gas and the self-gravitational pressure are given as
\begin{eqnarray}
P_\mathrm{ram} &=& \frac{F_\mathrm{drag}}{\pi a^2}\label{eq:Pram}\\
P_\mathrm{grav} &=& \frac{Gm^2}{\pi a^2\cdot a^2}\label{eq:Pgrav},
\end{eqnarray}
respectively.
Then, we take the maximum volume filling factor among Equation (\ref{eq:hitandstickphi}) and those when $P_\mathrm{comp}=P_\mathrm{ram}$ and $P_\mathrm{comp}=P_\mathrm{grav}$ for a given mass of dust aggregates.

\section{Results} \label{sec:results}

In this section, we show our results how large and fluffy dust aggregates are rotationally disrupted in protoplanetary disks.
First, we show steady-state spin period and tensile stress of rotating dust aggregates in Section \ref{subsec:period} and Section \ref{subsec:tensile}, respectively.
We assume two cases: compact case and porous case, in which volume filling factor of dust aggregates is fixed to be 0.5 and $10^{-3}$, respectively.
These volume filling factors correspond to the non-disrupted case and disrupted case, respectively.
Next, we show the mass and volume filling factor of rotationally disrupted dust aggregates as the fiducial case in Section \ref{subsec:disruption}.
Then, we investigate the dependence on the force-to-torque efficiency, orbital radius, and monomer radius in Section \ref{subsec:depend}.

\subsection{Steady-state Spin Period} \label{subsec:period}

\begin{figure*}
\plottwo{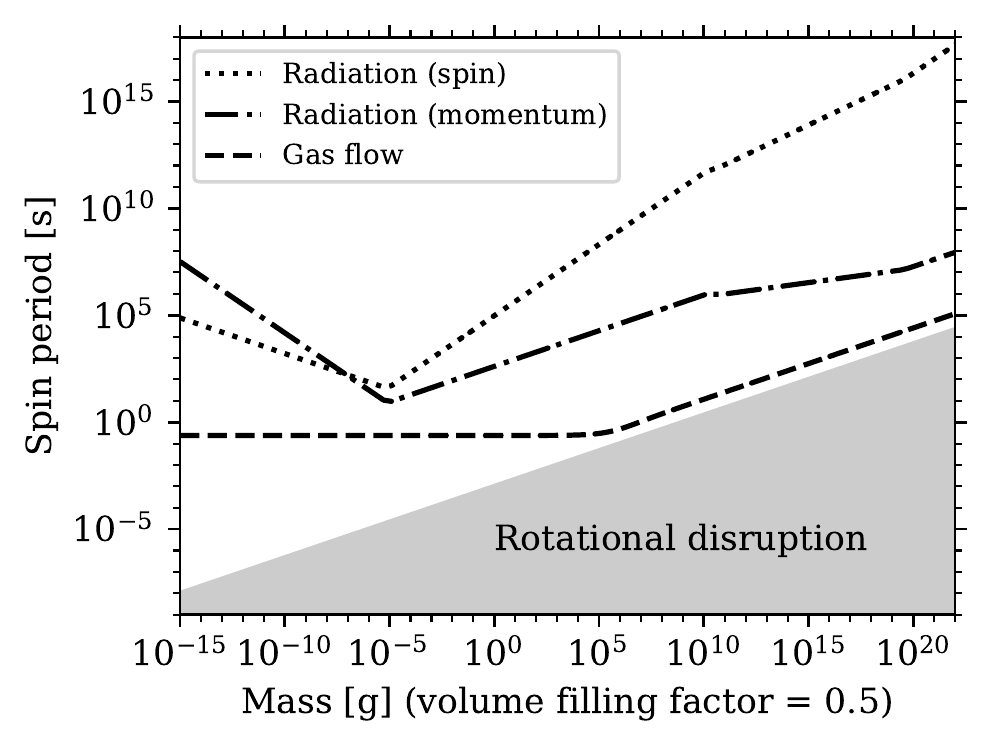}{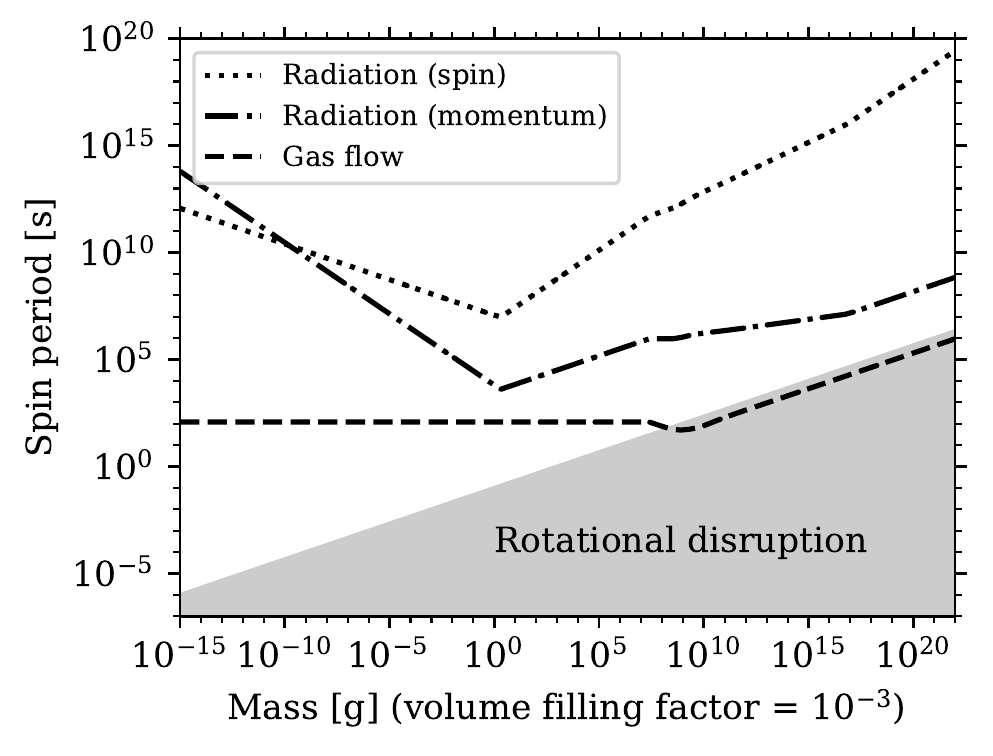}
\caption{Spin periods when $R=10\mathrm{\ au}$, $\gamma_\mathrm{p}=0.1$, $\gamma_\mathrm{ft}=0.1$, and $r_0=0.1\mathrm{\ \mu m}$ for compact dust aggregates (left; $\phi=0.5$) and porous dust aggregates (right; $\phi=10^{-3}$).
We assume three kinds of torques: radiative torque due to spin angular momentum of photons (dotted), radiative torque due to photon momentum (dash-dotted), and gas-flow torque (dashed).
The gray area represents where dust aggregates are rotationally disrupted.
\label{fig:spinperiod}}
\end{figure*}

We calculate the steady-state spin period $2\pi/\omega_\mathrm{c}$ using Equation (\ref{eq:constangvel}) for every spin-up torque when the volume filling factors are fixed to $\phi=0.5$ (compact) and $\phi=10^{-3}$ (porous) and plot it in Figure \ref{fig:spinperiod}.
We also show the rotational-disruption areas derived from the tensile strength (Equation (\ref{eq:tensilestrength})).
The other parameters are as follows: $R=10\mathrm{\ au}$, $\gamma_\mathrm{p}=0.1$, $\gamma_\mathrm{ft}=0.1$, and $r_0=0.1\mathrm{\ \mu m}$.

Figure \ref{fig:spinperiod} shows that the effect of radiative torques is much weaker than that of the gas-flow torque, which means that only the gas-flow torque is important in protoplanetary disks.
In protoplanetary disks, the gas density is much larger than that of the interstellar medium.
Therefore the spin-down torque due to the elastic collision of surrounding gas molecules is strong, which leads to the negligible effect of the radiative torques.

The dependence of each spin period on the mass is as follows.
The turn-over point of radiative torques around $m\sim10^{-5}$ g for $\phi=0.5$ and $m\sim1$ g for $\phi=10^{-3}$ correspond to $\lambda\sim1.8a\phi$, which is the boundary between two regimes of the radiative-torque efficiency (see Equation (\ref{eq:Qrad})).
The bend of gas-flow torque around $m\sim10^5$ g for $\phi=0.5$ and $m\sim10^9$ g for $\phi=10^{-3}$ corresponds to $\St\sim1$.
If the Stokes number is much smaller than unity ($\St\ll1$), the relative velocity between dust and gas has only the radial component (Equation (\ref{eq:vradial})), which means that $v\simeq v_r\simeq2\St\eta v_\mathrm{K}$.
On the other hand, if the Stokes number is much larger than unity ($\St\gg1$), the relative velocity has only the azimuthal component (Equation (\ref{eq:vazimuth})), which means that $v\simeq v_\varphi\simeq\eta v_\mathrm{K}$, where $\eta v_\mathrm{K}$ is constant in this paper.

{\tatsuuma{We analytically calculate the dependence of the steady-state angular velocity by assuming that the spin-up torque is the same as the gas-flow torque.
We also assume the Epstein regime for simplicity.
We substitute Equation (\ref{eq:Gammaupgas}), (\ref{eq:Fdrag}), (\ref{eq:stoppingtime}), (\ref{eq:vradial}), and (\ref{eq:vazimuth}) into Equation (\ref{eq:constangvel}), and then we obtain the dependence of the steady-state angular velocity written as}}
\begin{equation}
\omega_\mathrm{c}\propto\left\{
\begin{aligned}
& \phi R^{3/2}\gamma_\mathrm{ft} & (\St\ll1)\\
& m^{-1/3}\phi^{1/3}\gamma_\mathrm{ft} & (\St\gg1)
\end{aligned} \right..
\label{eq:omegadepend}
\end{equation}
Then, we find that the steady-state spin period ($\propto\omega_\mathrm{c}^{-1}$) does not depend on the mass when $\St\ll1$, while it is proportional to $m^{1/3}$ when $\St\gg1$ as shown in Figure \ref{fig:spinperiod}.

In the porous case ($\phi=10^{-3}$), dust aggregates with $m\gtrsim10^8$ g are rotationally disrupted by the gas flow, while the compact dust aggregates ($\phi=0.5$) are not disrupted.

\subsection{Tensile Stress due to Centrifugal Force} \label{subsec:tensile}

\begin{figure*}
\plottwo{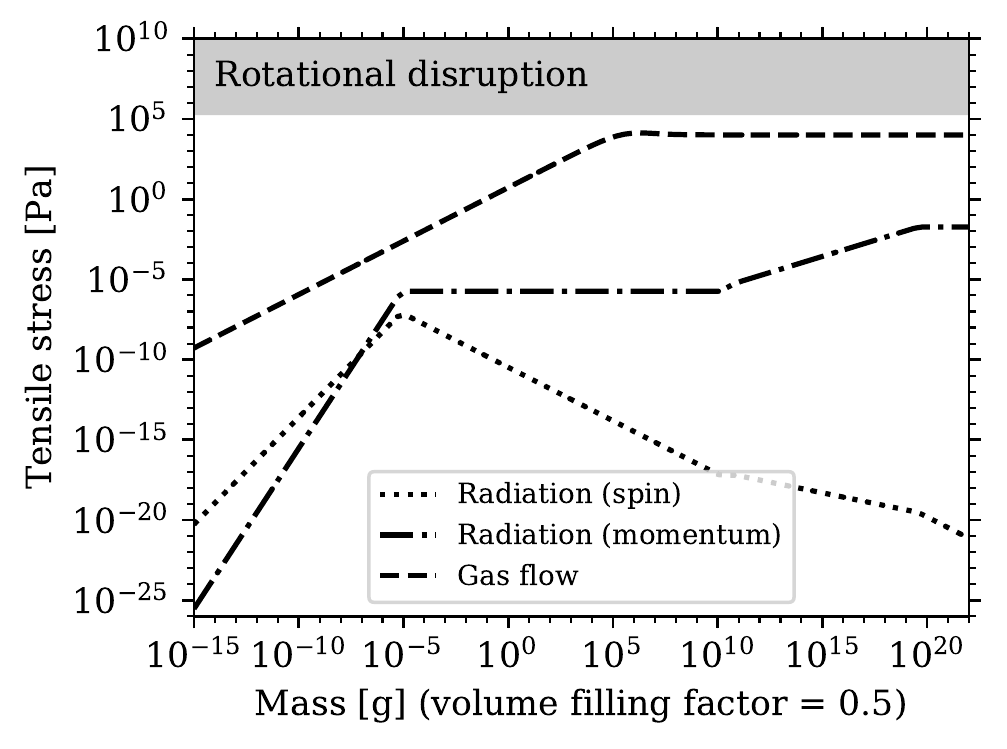}{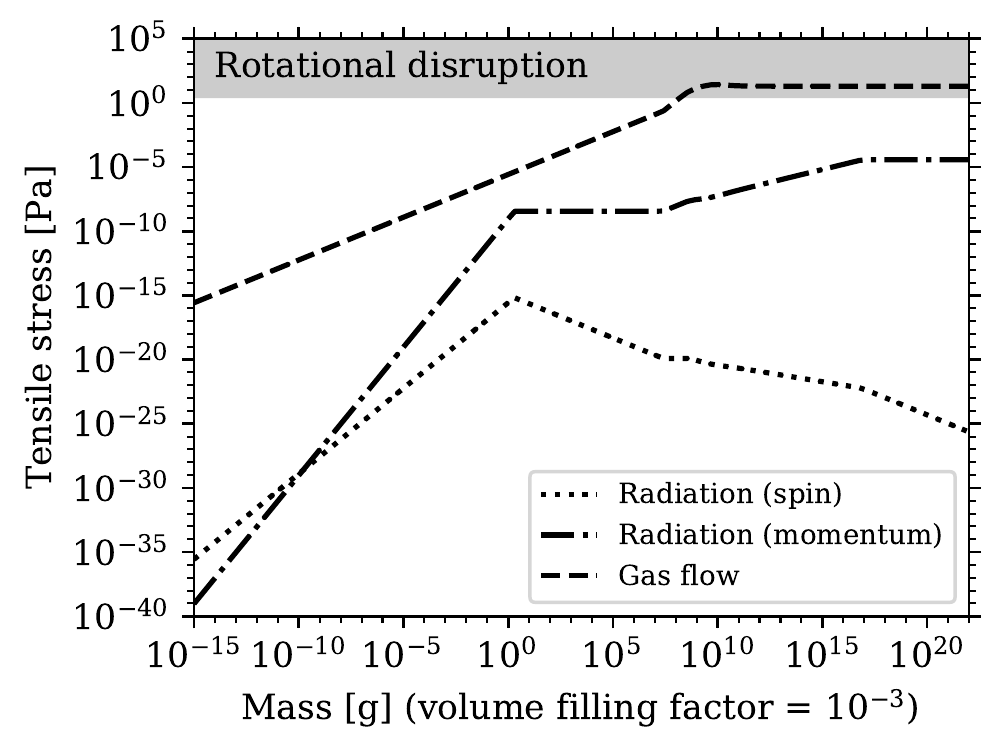}
\caption{Tensile stress when $R=10\mathrm{\ au}$, $\gamma_\mathrm{p}=0.1$, $\gamma_\mathrm{ft}=0.1$, and $r_0=0.1\mathrm{\ \mu m}$ for compact dust aggregates (left; $\phi=0.5$) and porous dust aggregates (right; $\phi=10^{-3}$).
The lines and the area are the same as Figure \ref{fig:spinperiod}.
\label{fig:tensile}}
\end{figure*}

We plot the tensile stress due to the centrifugal force using Equation (\ref{eq:stress}) for every spin-up torque when the volume filling factors are fixed to $\phi=0.5$ (compact) and $\phi=10^{-3}$ (porous) in Figure \ref{fig:tensile}.
The parameters are the same as Section \ref{subsec:period}: $R=10\mathrm{\ au}$, $\gamma_\mathrm{p}=0.1$, $\gamma_\mathrm{ft}=0.1$, and $r_0=0.1\mathrm{\ \mu m}$.
We also show the rotational-disruption area derived from the tensile strength of dust aggregates (Equation (\ref{eq:tensilestrength})), which does not depend on their mass.
The effect of radiative torques is also negligible because the tensile stress is proportional to the square of the steady-state angular velocity (see Equation (\ref{eq:stress})).

{\tatsuuma{We analytically calculate the dependence of the tensile stress by assuming that the spin-up torque is the same as the gas-flow torque.
We substitute Equation (\ref{eq:omegadepend}) into (\ref{eq:stress}), and then we obtain}}
\begin{equation}
S\propto\left\{
\begin{aligned}
& m^{2/3}\phi^{7/3} R^3\gamma_\mathrm{ft}^2 & (\St\ll1)\\
& \phi\gamma_\mathrm{ft}^2 & (\St\gg1)
\end{aligned} \right.,
\label{eq:stressdepend}
\end{equation}
which corresponds to the trend of Figure \ref{fig:tensile}.

\subsection{Disruption Mass \& Volume Filling Factor} \label{subsec:disruption}

\begin{figure}
\plotone{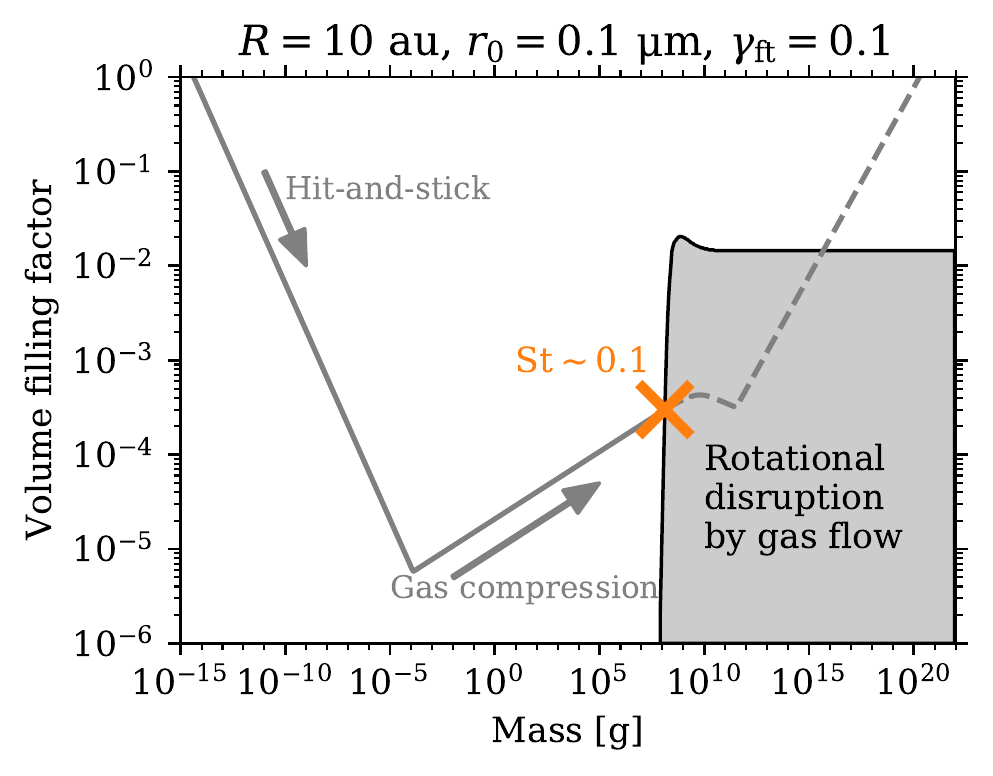}
\caption{Rotational disruption mass and volume filling factor of dust aggregates (gray area {\tatsuuma{with the black solid enclosing line}}) when $R=10\mathrm{\ au}$, $\gamma_\mathrm{p}=0.1$, $\gamma_\mathrm{ft}=0.1$, and $r_0=0.1\mathrm{\ \mu m}$.
Radiative torques are not strong enough to disrupt dust aggregates.
The dark-gray solid line represents the porosity evolution of dust aggregates, whose direction is annotated by the dark-gray arrow.
{\tatsuuma{This dark-gray solid line encounters the rotational disruption area when the Stokes number of dust aggregates is $\sim0.1$ (orange cross).
The dark-gray dashed line represents the growth of dust aggregates, which may not realize because of their rotational disruption.}}
\label{fig:fiducial}}
\end{figure}

We compare the calculated tensile stress due to the centrifugal force with the tensile strength of dust aggregates to determine whether they are rotationally disrupted.
We plot the disrupted mass and volume filling factor in Figure \ref{fig:fiducial}.
The other parameters are the same as Section \ref{subsec:period} and Section \ref{subsec:tensile}: $R=10\mathrm{\ au}$, $\gamma_\mathrm{p}=0.1$, $\gamma_\mathrm{ft}=0.1$, and $r_0=0.1\mathrm{\ \mu m}$.
Hereafter, we define this parameter set as the fiducial model and investigate the dependence on each parameter in Section \ref{subsec:depend}.

In the fiducial model, the dust aggregates with $m\gtrsim10^8$ g and $\phi\leq10^{-2}$ are rotationally disrupted by the gas flow.
Also, there is no dust aggregate rotationally disrupted by radiation.
We confirm that the contribution of radiative torque is negligible because gas-flow torque dominates for all dust aggregates even if they are small and compact.
See Appendix \ref{apsec:property} for further details, where we plot the spin period and tensile stress for every mass and volume filling factor of dust aggregates.

We also plot the dust porosity evolution in Figure \ref{fig:fiducial} and find that the dust aggregates with $\St\sim0.1$ are rotationally disrupted in their growth and compression processes {\tatsuuma{(See Appendix \ref{apsec:Stokes} for further details of the Stokes number)}}.
This corresponds to the dust aggregate with $m\sim10^8$ g, $\phi\sim3\times10^{-4}$, and thus $a\sim50$ m.
This dust aggregate rotates with $\omega_\mathrm{c}\sim4\times10^{-2}\mathrm{\ rad\ s^{-1}}$, which corresponds to the spin period $\sim3$ minutes and the velocity at its edge $\sim2\mathrm{\ m\ s^{-1}}$.
The tensile stress due to this rotation is $\sim0.3$ Pa.
These properties of rotationally disrupted dust aggregates are also listed in Table \ref{tab:dustproperty}.

\begin{table*}
\centering
\caption{Properties of rotationally disrupted dust aggregates during their porosity evolution.
We show all the models treated in this paper.
ND represents Non-Disruption, which means that dust aggregates are not rotationally disrupted during their porosity evolution.
} \label{tab:dustproperty}
\begin{tabular}{ccc|cccccccc}
\tablewidth{0pt}
\hline
\hline
$R$ & $\gamma_\mathrm{ft}$ & $r_0$ & $m$ & $\phi$ & $a$ & $\St$ & $\omega_\mathrm{c}$ & $2\pi/\omega_\mathrm{c}$ & $a\omega_\mathrm{c}$ & $S$\\
\hline
10 au & 0.1 & 0.1 $\mathrm{\mu m}$ & $10^8$ g & $3\times10^{-4}$ & 50 m & 0.1 & $4\times10^{-2}\mathrm{\ rad\ s^{-1}}$ & 3 min & 2 $\mathrm{m\ s^{-1}}$ & 0.3 Pa\\
10 au & 0.01 & 0.1 $\mathrm{\mu m}$ & ND & ND & ND & ND & ND & ND & ND & ND\\
10 au & 0.02 & 0.1 $\mathrm{\mu m}$ & ND & ND & ND & ND & ND & ND & ND & ND\\
10 au & 0.03 & 0.1 $\mathrm{\mu m}$ & $2\times10^9$ g & $4\times10^{-4}$ & 100 m & 0.5 & $2\times10^{-2}\mathrm{\ rad\ s^{-1}}$ & 5 min & 2 $\mathrm{m\ s^{-1}}$ & 0.5 Pa\\
5 au & 0.1 & 0.1 $\mathrm{\mu m}$ & $4\times10^7$ g & $4\times10^{-4}$ & 30 m & 0.1 & $7\times10^{-2}\mathrm{\ rad\ s^{-1}}$ & 1 min & 2 $\mathrm{m\ s^{-1}}$ & 0.5 Pa\\
50 au & 0.1 & 0.1 $\mathrm{\mu m}$ & $6\times10^6$ g & $7\times10^{-5}$ & 30 m & 0.06 & $4\times10^{-2}\mathrm{\ rad\ s^{-1}}$ & 3 min & 1 $\mathrm{m\ s^{-1}}$ & 0.02 Pa\\
10 au & 0.1 & 1.0 $\mathrm{\mu m}$ & $2\times10^7$ g & $2\times10^{-3}$ & 10 m & 0.07 & $9\times10^{-2}\mathrm{\ rad\ s^{-1}}$ & 1 min & 1 $\mathrm{m\ s^{-1}}$ & 0.6 Pa\\
\hline
\end{tabular}
\end{table*}

\subsection{Parameter Dependence} \label{subsec:depend}

In this section, we investigate how mass and volume filling factor of the disrupted dust aggregates depend on the force-to-torque efficiency, orbital radius, and monomer radius in Section \ref{subsubsec:dependft}, Section \ref{subsubsec:dependorbit}, and Section \ref{subsubsec:dependmonomer}, respectively.
In the fiducial model, we find that the gas-flow torque is much stronger than the radiative torques and the dust aggregates with $m\gtrsim10^8$ g and $\phi\leq10^{-2}$ are rotationally disrupted by the gas flow.
We investigate how this result changes for different models.
We also show the spin period and tensile stress for every model in Appendix \ref{apsubsec:period} and Appendix \ref{apsubsec:stress}, respectively.

\subsubsection{Force-to-torque Efficiency} \label{subsubsec:dependft}

\begin{figure}
\plotone{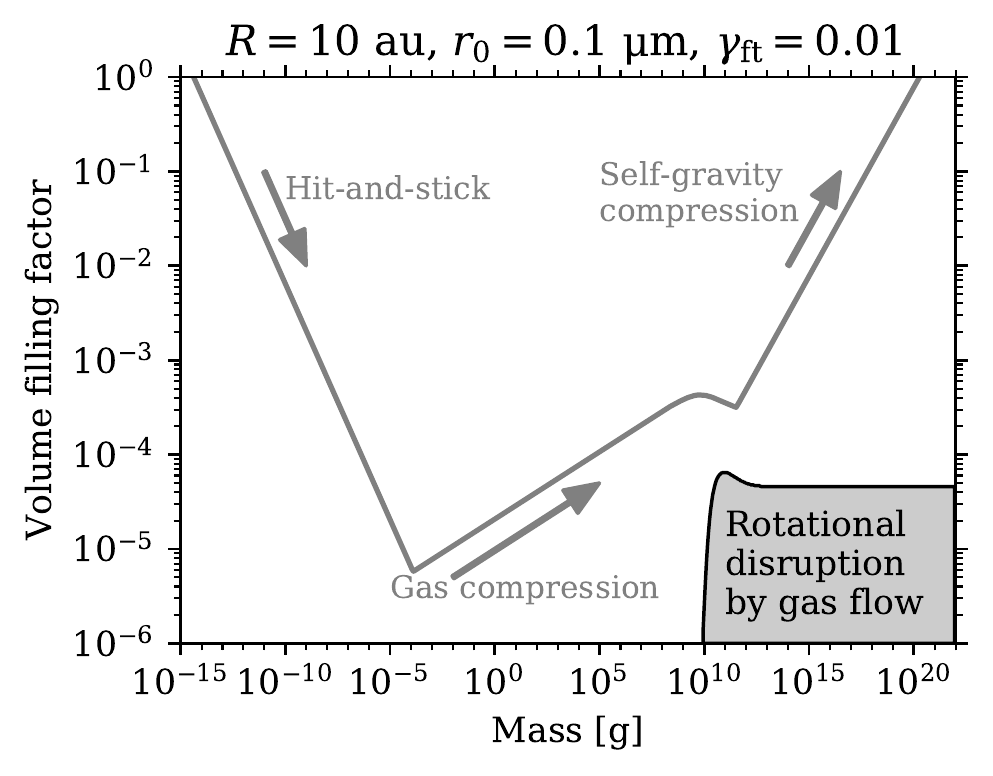}
\caption{Rotational disruption mass and volume filling factor of dust aggregates (gray area {\tatsuuma{with the black solid enclosing line}}) when $R=10\mathrm{\ au}$, $\gamma_\mathrm{p}=0.1$, $\gamma_\mathrm{ft}=0.01$, and $r_0=0.1\mathrm{\ \mu m}$.
The lines are the same as Figure \ref{fig:fiducial}.
\label{fig:forcetorque}}
\end{figure}

To investigate the dependence on the force-to-torque efficiency, we plot the mass and volume filling factor of rotationally disrupted dust aggregates with $\gamma_\mathrm{ft}=0.01$ in Figure \ref{fig:forcetorque}.
{\mtatsuuma{The force-to-torque efficiency is introduced in Section \ref{subsubsec:gastorque} as the rate of the drag force converted to the torque, interpreted in Section \ref{subsec:interpangvel}, and discussed the effect on the planetesimal formation in Section \ref{subsec:planetform}.}}
The other parameters are the same as the fiducial model: $R=10\mathrm{\ au}$, $\gamma_\mathrm{p}=0.1$, and $r_0=0.1\mathrm{\ \mu m}$.

The rotational-disruption area, where $m\gtrsim10^{10}$ g and $\phi\lesssim4\times10^{-5}$, is smaller than that of the fiducial model (Figure \ref{fig:fiducial}) because the tensile stress is proportional to $\gamma_\mathrm{ft}^2$ (Equation (\ref{eq:stressdepend})).

In this case, growing dust aggregates according to their porosity evolution are not rotationally disrupted.
This means that the force-to-torque efficiency is a key parameter of the rotational disruption.
We confirm that dust aggregates are rotationally disrupted when $\gamma_\mathrm{ft}=0.03$, while they are not rotationally disrupted when $\gamma_\mathrm{ft}=0.02$ in the fiducial model.
The properties of rotationally disrupted dust aggregates when $\gamma_\mathrm{ft}=0.02$ and 0.03 are listed in Table \ref{tab:dustproperty}.

\subsubsection{Orbital Radius} \label{subsubsec:dependorbit}

\begin{figure*}
\plottwo{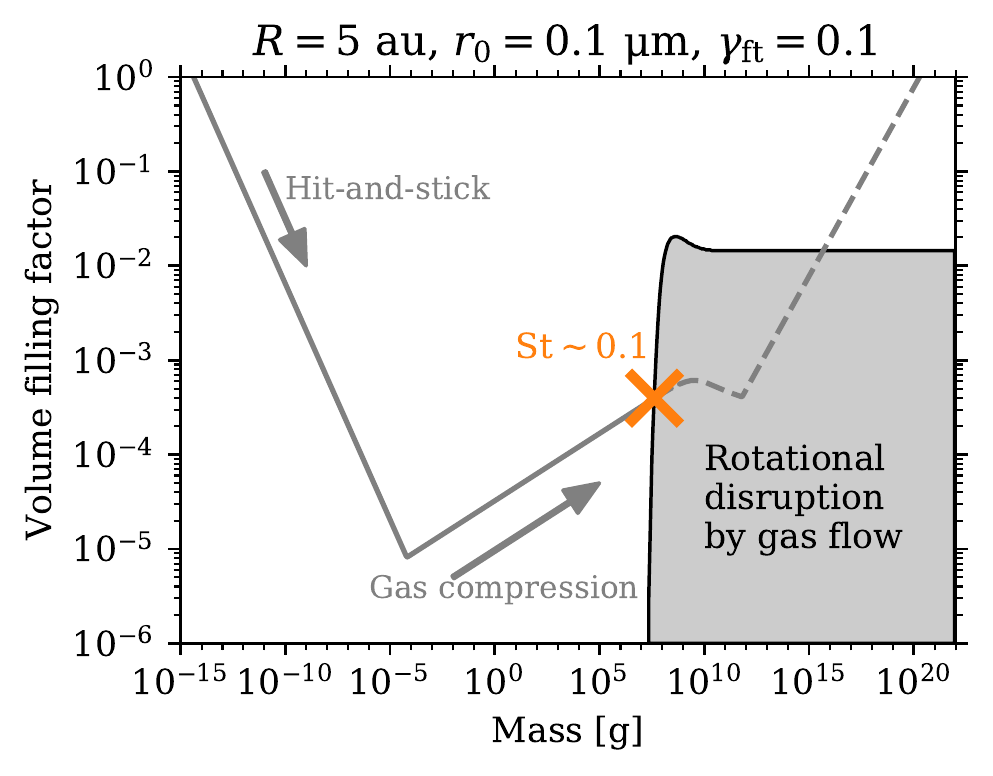}{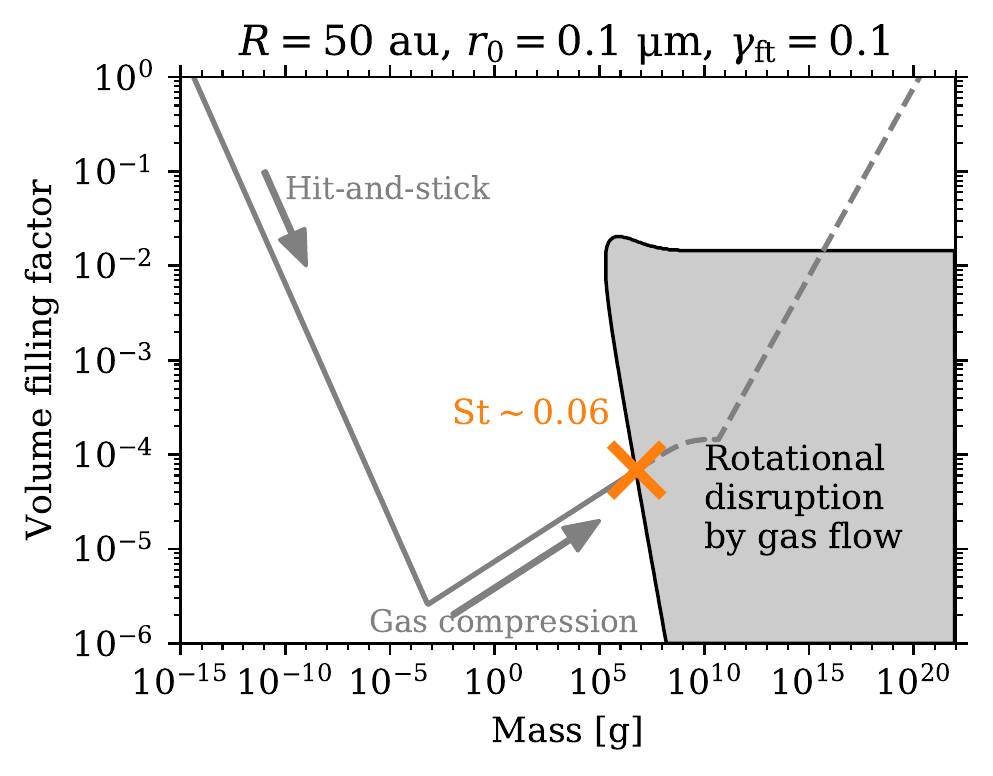}
\caption{Rotational disruption mass and volume filling factor of dust aggregates (gray area {\tatsuuma{with the black solid enclosing line}}) when $\gamma_\mathrm{p}=0.1$, $\gamma_\mathrm{ft}=0.1$, and $r_0=0.1\mathrm{\ \mu m}$.
The orbital radii are $R=5\mathrm{\ au}$ (left) and $R=50\mathrm{\ au}$ (right).
The lines are the same as Figure \ref{fig:fiducial}.
\label{fig:orbitradius}}
\end{figure*}

We plot the rotational-disruption area with different orbital radius in Figure \ref{fig:orbitradius}.
The orbital radii are 5 au and 50 au and the other parameters are the same as the fiducial model: $\gamma_\mathrm{p}=0.1$, $\gamma_\mathrm{ft}=0.1$, and $r_0=0.1\mathrm{\ \mu m}$.

The mass of disrupted dust aggregates is $m\gtrsim10^7$ g for 5 au and $m\gtrsim10^5$ g for 50 au, respectively, while the volume filling factor is the same.
This difference in the disrupted mass arises because the tensile stress is proportional to $R^3$ when $\St\ll1$ (Equation (\ref{eq:stressdepend})).
In contrast, there is no difference in the disrupted volume filling factor because the tensile stress does not depend on the orbital radius when $\St\gg1$ (Equation (\ref{eq:stressdepend})).

In both cases, growing dust aggregates according to their porosity evolution are rotationally disrupted.
The properties of these dust aggregates are listed in Table \ref{tab:dustproperty}.

\subsubsection{Monomer Radius} \label{subsubsec:dependmonomer}

\begin{figure}
\plotone{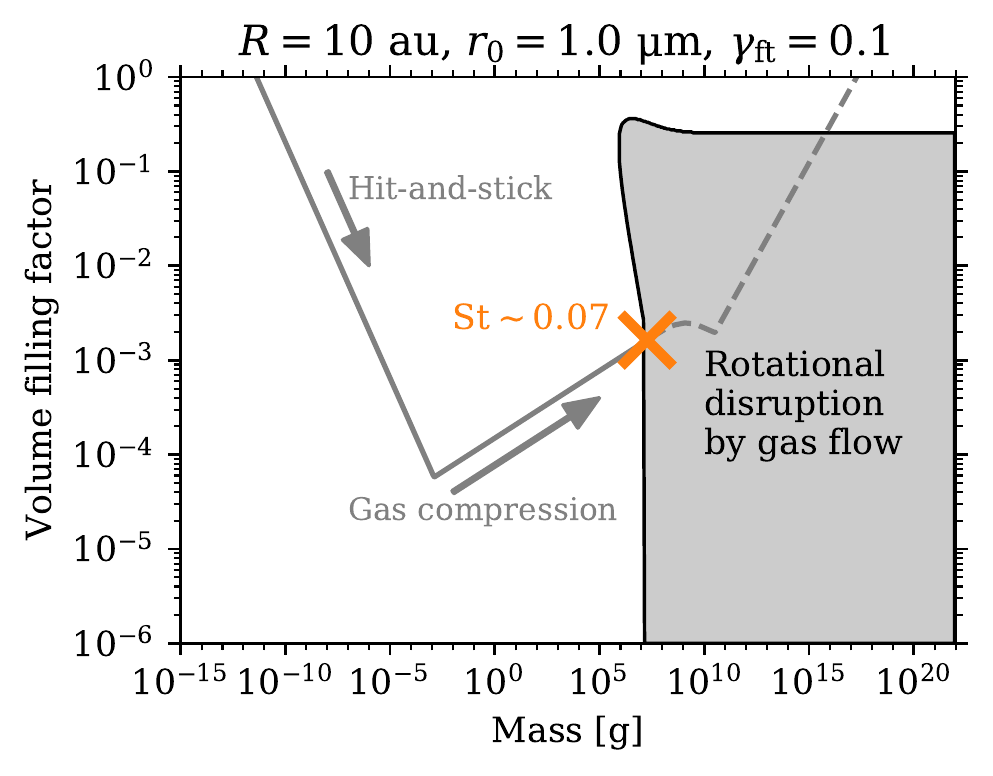}
\caption{Rotational disruption mass and volume filling factor of dust aggregates (gray area {\tatsuuma{with the black solid enclosing line}}) when $R=10\mathrm{\ au}$, $\gamma_\mathrm{p}=0.1$, $\gamma_\mathrm{ft}=0.1$, and $r_0=1.0\mathrm{\ \mu m}$.
The lines are the same as Figure \ref{fig:fiducial}.
\label{fig:monomerradius}}
\end{figure}

We investigate the dependence on the monomer radius because it may be larger than sub-micrometer \citep[e.g.,][]{Tatsuuma2019}.
In Figure \ref{fig:monomerradius}, we plot the rotational-disruption area with $r_0=1.0\mathrm{\ \mu m}$.
The other parameters are the same as the fiducial model: $R=10\mathrm{\ au}$, $\gamma_\mathrm{p}=0.1$, and $\gamma_\mathrm{ft}=0.1$.

The rotational-disruption area, where $m\gtrsim10^7$ g and $\phi\lesssim0.2$, is larger than that of the fiducial model (Figure \ref{fig:fiducial}).
The gas-flow torque does not depend on the monomer radius, while the tensile strength is inversely proportional to it (Equation (\ref{eq:tensilestrength})).
Thus, dust aggregates are more fragile in this case than the fiducial model.

In this case, growing dust aggregates according to their porosity evolution are rotationally disrupted and their properties are listed in Table \ref{tab:dustproperty}.
Although the properties are different among all models, there is a common tendency that highly porous dust aggregates whose radii are several tens meters are rotationally disrupted.

\section{Discussion} \label{sec:discussion}

In this section, we discuss our results for a detailed understanding of rotational disruption.
We discuss the steady-state angular velocity of rotating dust aggregates by gas flow in protoplanetary disks in Section \ref{subsec:interpangvel}.
To determine whether dust aggregates are rotationally disrupted, we estimate timescales of dust growth, rotation, and sound-crossing inside dust aggregates in Section \ref{subsec:timescales}.
We find that the force-to-torque efficiency is a key parameter in Section \ref{subsubsec:dependft}, and thus we discuss it by using the toy model introduced by \citet{Lazarian2007MA} in Section \ref{subsec:whatisft?}.
Although we assume that there is no turbulence in this paper, strong turbulence may prevent dust aggregates from rotational disruption.
We discuss the condition of turbulence by using $\alpha$ parameter in Section \ref{subsec:turbulence}.

Also, we show some implications for observation of protoplanetary disks by using the absorption mass opacity of porous dust aggregates \citep{Kataoka2014} and the planetesimal-formation mechanism in Section \ref{subsec:diskobs} and Section \ref{subsec:planetform}, respectively.

\subsection{Steady-state Angular Velocity due to Gas Flow} \label{subsec:interpangvel}

Our results show that dust aggregates are rotationally disrupted by gas flow in protoplanetary disks, which leads to the easy interpretation of the steady-state angular velocity.
When $\Gamma_\mathrm{up}=\Gamma_\mathrm{up,gas}$, we obtain the simple description of $\omega_\mathrm{c}$ from Equation (\ref{eq:constangvel}) as
\begin{equation}
\omega_\mathrm{c} = \frac{5}{3}\frac{v\gamma_\mathrm{ft}}{a}\sim\frac{v}{a}\gamma_\mathrm{ft}.
\label{eq:constangvelgas}
\end{equation}

We can interpret Equation (\ref{eq:constangvelgas}) as follows.
If the force-to-torque efficiency is $\gamma_\mathrm{ft}=1$, $\omega_\mathrm{c}\sim v/a$, which means that the rotational velocity at the edge of dust aggregates is the same as the relative velocity between dust and gas.
In other words, we can define $\gamma_\mathrm{ft}$ as the difference between the relative velocity and the rotational velocity at the edge.

\subsection{Timescales} \label{subsec:timescales}

For a dust aggregate to be rotationally disrupted, it has to reach a rotational steady-state and also the tensile stress has to travel through it before its growth.
In other words, the timescale of dust growth has to be larger than that of rotation and sound-crossing time.
For the fiducial model in Section \ref{sec:results}, the lightest dust aggregates have $\St\sim0.1$ or $m\sim10^8$ g.
Thus, we assume the dust aggregates with $\St\sim0.1$ or $m\sim10^8$ g and estimate timescales of dust growth, rotation, and sound-crossing.

The timescale of dust growth is as follows \citep[see][]{Okuzumi2012}.
This timescale is defined as
\begin{equation}
t_\mathrm{grow} = \frac{m}{\dot{m}}=\frac{m}{\pi a^2v_\mathrm{t}\rho_\mathrm{dust}},
\label{eq:tgrowfirst}
\end{equation}
where $v_\mathrm{t}\sim\sqrt{\alpha\St}c_\mathrm{s}$ is the turbulence-driven relative velocity between dust aggregates \citep{Ormel2007}, $\alpha$ is the turbulent parameter, $\rho_\mathrm{dust}=\Sigma_\mathrm{dust}/(\sqrt{2\pi}h_\mathrm{dust})$ is the dust density, $\Sigma_\mathrm{dust}$ is the dust surface density, and $h_\mathrm{dust}=\sqrt{\alpha/\St}h_\mathrm{gas}$ is the dust scale height.
Then, Equation (\ref{eq:tgrowfirst}) can be written as
\begin{equation}
t_\mathrm{grow} \sim\frac{\sqrt{2\pi}mh_\mathrm{gas}}{\pi a^2\St c_\mathrm{s}\Sigma_\mathrm{dust}}.
\label{eq:tgrowsecond}
\end{equation}
Here, we assume the Epstein-drag regime for simplicity, and then
\begin{equation}
t_\mathrm{grow} \sim 30\left(\frac{\Sigma_\mathrm{gas}/\Sigma_\mathrm{dust}}{100}\right)t_\mathrm{K},
\label{eq:tgrowfinal}
\end{equation}
where $t_\mathrm{K}=2\pi/\Omega_\mathrm{K}$, is obtained.

The timescale of rotation is the time required for dust aggregates to spin-up from $\omega=0$ to $\omega=\omega_\mathrm{c}$.
If there is only spin-up torque due to the gas flow, the timescale of rotational disruption is given as
\begin{equation}
t_\mathrm{rot} = \frac{\omega_\mathrm{c}}{\mathrm{d}\omega/\mathrm{d}t} = \frac{\omega_\mathrm{c}}{\Gamma_\mathrm{up,gas}/I} = \frac{3}{5}\frac{a\omega_\mathrm{c}t_\mathrm{s}}{v\gamma_\mathrm{ft}}.
\label{eq:trotfirst}
\end{equation}
By substituting Equation (\ref{eq:constangvelgas}), we obtain
\begin{equation}
t_\mathrm{rot}\sim t_\mathrm{s}\sim0.02\left(\frac{\St}{0.1}\right)t_\mathrm{K}.
\label{eq:trotfinal}
\end{equation}

The timescale of sound-crossing inside dust aggregates is written by using the effective sound velocity as \citep{Tatsuuma2019}
\begin{eqnarray}
t_\mathrm{cross} &=& \frac{a}{c_\mathrm{s,eff}} \sim \frac{a}{2.2\times10^3\phi\mathrm{\ cm\ s^{-1}}}\nonumber\\
&\sim& 10^7\left(\frac{m}{10^8\mathrm{\ g}}\right)^{1/3}\left(\frac{\phi}{10^{-6}}\right)^{-4/3}\mathrm{\ s},
\label{eq:tseff}
\end{eqnarray}
which means that the sound-crossing timescale is shorter than one year even for the extremely porous case.

By comparing Equation (\ref{eq:tgrowfinal}), Equation (\ref{eq:trotfinal}), and Equation (\ref{eq:tseff}), we find that $t_\mathrm{grow}\gg t_\mathrm{rot}$ for typical gas-to-dust ratio and $t_\mathrm{grow}\gg t_\mathrm{cross}$ because $t_\mathrm{K}$ is larger than one year where $R>1$ au.
Therefore, dust aggregates can be rotationally disrupted before they grow.

However, it is unclear whether $t_\mathrm{rot}$ is larger than $t_\mathrm{cross}$.
We can treat the dust aggregate as a rigid body when $t_\mathrm{rot}\gg t_\mathrm{cross}$.
When $t_\mathrm{rot}\ll t_\mathrm{cross}$, the shear movement inside the dust aggregate may appear, which means that we have to use its shear strength.
The shear strength of dust aggregates for silicate was investigated by \citet{Seizinger2013} while that for ice is unclear.

\subsection{Force-to-torque Efficiency and a Toy Model} \label{subsec:whatisft?}

In this section, we discuss the corresponding expression of force-to-torque efficiency $\gamma_\mathrm{ft}$ to the model introduced by \citet{Lazarian2007MA}.
They used the toy model, which is a grain with a reflecting mirror attached to it by a pole.
We assume that the length of the pole is the same as the radius of dust aggregates $a$.
Then, the torque in their model can be written as
\begin{equation}
\Gamma_\mathrm{LH07} = \frac{m_\mathrm{H}nv_\mathrm{th}^2Aa}{2}Q_\mathrm{\Gamma LH07},
\label{eq:gammaLH07}
\end{equation}
where $m_\mathrm{H}$ is the mass of a hydrogen atom, $n$ is its number density, $A$ is the area of the mirror, and $Q_\mathrm{\Gamma LH07}$ is the torque efficiency in their model.
Here, we simply assume that $m_\mathrm{H}n\sim\rho_\mathrm{gas}$.
The gas-flow torque (Equation (\ref{eq:Gammaupgas})) can be written as
\begin{equation}
\Gamma_\mathrm{up,gas} = \frac{8}{9}\pi\rho_\mathrm{gas}v_\mathrm{th}va^3\gamma_\mathrm{ft},
\label{eq:gammagasEpstein}
\end{equation}
where we assume the Epstein-drag regime.
Then, we obtain an expression for $\gamma_\mathrm{ft}$ as
\begin{equation}
\gamma_\mathrm{ft} \sim 0.3\left(\frac{v}{c_\mathrm{s}}\right)^{-1}\left(\frac{A}{a^2}\right)Q_\mathrm{\Gamma LH07},
\label{eq:toymodel}
\end{equation}
where the relative velocity between dust and gas $v$ is subsonic, for example, $v/c_\mathrm{s}=0.02$ when $R=10$ au and $\St=0.1$, the area of the mirror is smaller than $a^2$, for example, $A\sim(0.1a)^2$, and the torque efficiency $Q_\mathrm{\Gamma LH07}$ is about unity \citep{Lazarian2007MA}.
Then, we discuss the possible value of the force-to-torque efficiency as
\begin{equation}
\gamma_\mathrm{ft} \lesssim 0.3\left(\frac{1}{0.02}\right)(0.01)1\sim0.15,
\label{eq:toymodelex}
\end{equation}
which is consistent with the assumed value in this paper.

\subsection{Strength of Turbulence} \label{subsec:turbulence}

{\tatsuuma{If the turbulence is too strong, are dust aggregates rotationally disrupted?
Our assumption of weak turbulence would be incorrect in this case.
However, whether dust aggregates are rotationally disrupted depends on the eddy turn-over timescale of turbulence, the timescale of rotational disruption $t_\mathrm{rot}$ (Equation (\ref{eq:trotfinal})), and the timescale of sound-crossing $t_\mathrm{cross}$ (Equation (\ref{eq:tseff})).
They can be rotationally disrupted by torque originated from the turbulence if the eddy turn-over timescale is longer than $t_\mathrm{rot}\sim t_\mathrm{s}$ (Equation (\ref{eq:trotfinal})).
Also, they may be rotationally disrupted by strong turbulence if the eddy turn-over timescale is longer than $t_\mathrm{cross}$ even if the eddy turn-over timescale is shorter than $t_\mathrm{rot}$.
It is unclear which timescale is the longest, and thus the effect of strong turbulence is unknown.}}

At least, we can discuss the validity of the assumption of weak turbulence.
The turbulence would not affect the relative velocity between dust aggregates and gas $v$ if turbulent velocity is smaller than $v$.
The turbulent velocity can be described as $v_\mathrm{t}\sim\sqrt{\alpha\St}c_\mathrm{s}$ \citep{Ormel2007}.
Thus, we obtain the condition of $\alpha$ for rotational disruption as
\begin{equation}
\alpha<4\times10^{-3}\left(\frac{v/c_\mathrm{s}}{0.02}\right)^2\left(\frac{\St}{0.1}\right)^{-1}.
\label{eq:upperturb}
\end{equation}
This condition is consistent with some observations of molecular lines, polarization, and continuum, which infer weak turbulence \citep[e.g.,][]{Flaherty2015,Flaherty2017,Flaherty2018,Pinte2016,Teague2016,Ohashi2019}.

\subsection{{\misako{Internal Alignment}}} \label{subsec:internal}

{\misako{In this section, we discuss the effect of the internal alignment of dust aggregates.
The internal alignment is an alignment of the spin axis with the minor axis of dust grains \citep[e.g.,][]{Purcell1979}.
When dust grains begin to rotate, the spin axis is not parallel to their minor axis.
This precession would be damped, which would affect the grain alignment.}}

{\mtatsuuma{The polarized emission due to dust alignment in protoplanetary disks has been observed at radio wavelength \citep[e.g.,][]{Kataoka2017,Stephens2017}, which means that some polarization observations show that dust grains' precession motion would be damped.
Although the effect of the internal alignment of porous dust aggregates in the context of planet formation is still under debate, this paper focuses on the possibility of the rotational disruption in the simple model as the first step.}}

\subsection{Implication for Disk Observations} \label{subsec:diskobs}

If dust aggregates are rotationally disrupted, are they observable?
Dust aggregates at disk midplane emit thermal emission, which can be observed at radio wavelength.
Their observability is determined by their absorption mass opacity, which can be characterized by $a\phi$ \citep{Kataoka2014}.
Table \ref{tab:dustproperty} shows that $a\phi$ of the dust aggregates which are rotationally disrupted has a range between several millimeters and several centimeters.
This corresponds to the maximum size determined by the observed spectral index of protoplanetary disks \citep[e.g.,][for a review]{Testi2014}.

However, some polarization observations show that the maximum grain size can be around one hundred micrometer \citep[e.g.,][]{Kataoka2016HLTau,Kataoka2016HD,Kataoka2017}.
Also, it is pointed out that highly porous dust aggregates cannot explain polarization observations{\fourthcomment{, while moderate porosity ($\lesssim 0.7$) can be a solution}} \citep[e.g.,][]{Kirchschlager2019,Tazaki2019}.
The observed grain size and porosity are still under debate.

\subsection{Implication for Planetesimal Formation} \label{subsec:planetform}

How can planetesimals form if the dust growth is halted by rotational disruption?
To avoid rotational disruption, the compression mechanism without coagulation is needed because compact dust aggregates would not be rotationally disrupted.
Also, some ``lucky'' dust aggregates whose force-to-torque efficiencies are extremely small can grow without experiencing rotational disruption, and then they feed on other dust aggregates whose growth is halted because of rotational disruption \citep[e.g.,][]{Windmark2012}.
{\mtatsuuma{This means that both rotationally disrupted and non-disrupted dust aggregates can contribute to the planetesimal formation.}}
It is unclear how small the force-to-torque efficiency of dust aggregates is{\mtatsuuma{, which is beyond this paper}}.

\section{Conclusion} \label{sec:conclusions}

We theoretically calculated whether {\fourthcomment{irregularly-shaped}} porous dust aggregates can be disrupted by their spinning motion during their growth {\fourthcomment{at the midplane}} in protoplanetary disks as an application of the rotational disruption of dust grains in the interstellar space \citep[e.g.,][]{Hoang2019ApJ,Hoang2019NatAs}.
We assumed radiative torque \citep[e.g.,][]{Draine1996} and gas-flow torque as driving sources of the spinning motion.
We calculated the steady-state rotational velocity of the dust aggregates by assuming that their spinning motion is damped by the surrounding gas and they reach a steady-state rigid rotation.
After obtaining the tensile stress due to the centrifugal force, we compared it with the tensile strength of dust aggregates \citep{Tatsuuma2019}.
Our main findings of the rotational disruption of {\fourthcomment{irregularly-shaped}} porous dust aggregates are as follows.

\begin{enumerate}

\item Radiative torque on dust aggregates is much weaker than gas-flow torque {\fourthcomment{at the midplane}} in protoplanetary disks, and thus the radiative torque is negligible.

\item If dust aggregates are compact, where their volume filling factor is close to unity, they are not rotationally disrupted.

\item In our fiducial model, the dust aggregates whose mass is larger than $\sim10^8$ g and volume filling factor is smaller than $\sim10^{-2}$ are rotationally disrupted by the gas-flow torque.
The fiducial parameters are as follows: the orbital radius $R=10$ au, the force-to-torque efficiency of dust aggregates $\gamma_\mathrm{ft}=0.1$, and the radius of icy monomers $r_0=0.1\mathrm{\ \mu m}$.

\item If we assume the dust porosity evolution \citep{Kataoka2013L}, dust aggregates whose Stokes number is $\sim0.1$ can be rotationally disrupted in their growth and compression process.
This corresponds to the dust aggregates whose radius is around several tens meters.

\item Whether dust aggregates are rotationally disrupted depends on their force-to-torque efficiency {\fourthcomment{$\gamma_\mathrm{ft}$}}, which is not fully investigated.
{\fourthcomment{In our fiducial model, dust aggregates are not rotationally disrupted when $\gamma_\mathrm{ft}\leq0.02$.}}

\end{enumerate}

\acknowledgments

We thank Ryo Tazaki and Hidekazu Tanaka for the insightful discussions.
This work was supported by JSPS KAKENHI Grant Number 19J20351, 18K13590, and 19H05088.

\appendix

\section{Rotational Property for Every Dust Mass \& Volume Filling Factor} \label{apsec:property}

In this section, we show the spin period and tensile stress for every mass and volume filling factor of dust aggregates for every model in this paper.
We find that the gas-flow torque is dominant for all dust aggregates even if they are small and compact.
The spin period and tensile stress is shown in Section \ref{apsubsec:period} and Section \ref{apsubsec:stress}, respectively.

\subsection{Spin Period} \label{apsubsec:period}

\begin{figure}
\plotone{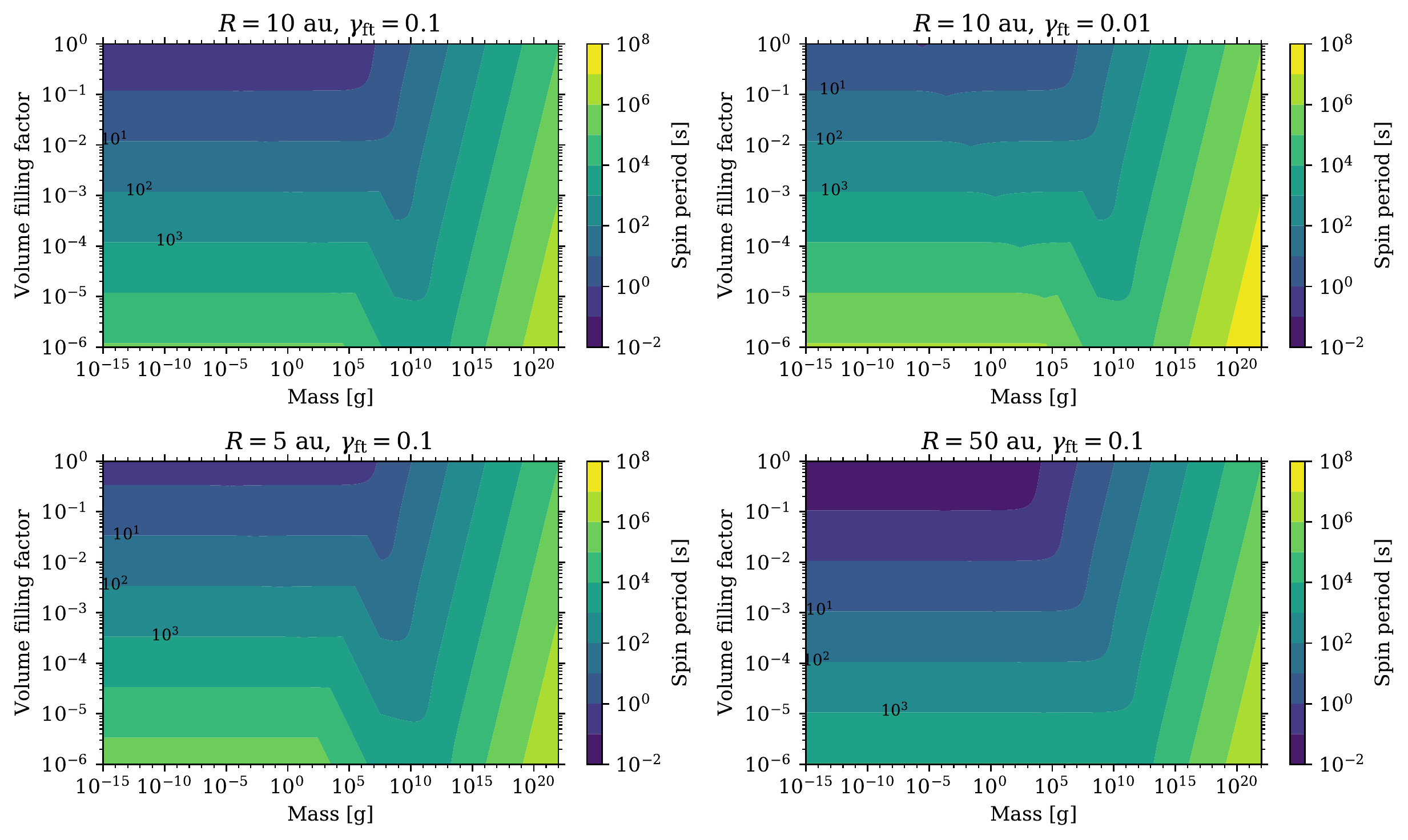}
\caption{Spin period (color) when $R=10$ au and $\gamma_\mathrm{ft}=0.1$ (upper left), $R=10$ au and $\gamma_\mathrm{ft}=0.01$ (upper right), $R=5$ au and $\gamma_\mathrm{ft}=0.1$ (lower left), and $R=50$ au and $\gamma_\mathrm{ft}=0.1$ (lower right).
The other parameter is the same as the fiducial model: $\gamma_\mathrm{p}=0.1$.
\label{fig:periodall}}
\end{figure}

We plot the spin period $2\pi/\omega_\mathrm{c}$ in Figure \ref{fig:periodall} by assuming three kinds of spin-up torques: the radiative torques due to spin angular momentum of photons and photon momentum and the gas-flow torque.
We find that the gas-flow torque is dominant for every mass and volume filling factor of dust aggregates.

\subsection{Tensile Stress} \label{apsubsec:stress}

\begin{figure}
\plotone{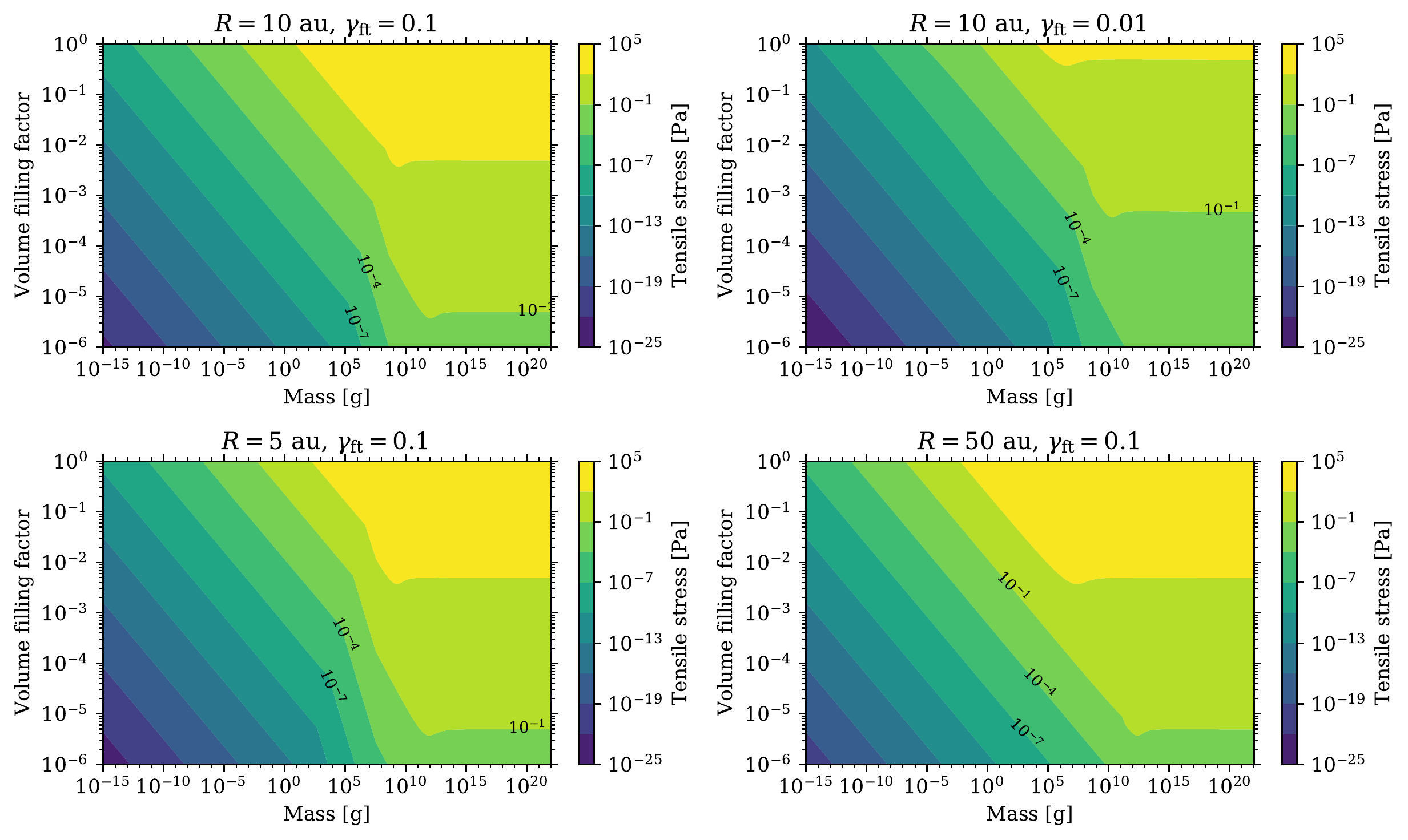}
\caption{Tensile stress (color) when $R=10$ au and $\gamma_\mathrm{ft}=0.1$ (upper left), $R=10$ au and $\gamma_\mathrm{ft}=0.01$ (upper right), $R=5$ au and $\gamma_\mathrm{ft}=0.1$ (lower left), and $R=50$ au and $\gamma_\mathrm{ft}=0.1$ (lower right).
The other parameter is the same as the fiducial model: $\gamma_\mathrm{p}=0.1$.
\label{fig:stressall}}
\end{figure}

We also plot the tensile stress $S$ in Figure \ref{fig:stressall} using the derived steady-state angular velocity $\omega_\mathrm{c}$ (see Equation (\ref{eq:stress})).
We find that the tensile stress is almost determined by the gas-flow torque for all parameters in this paper.

\section{{\tatsuuma{Stokes Number of Porous Dust Aggregates}}} \label{apsec:Stokes}

\begin{figure}
\plotone{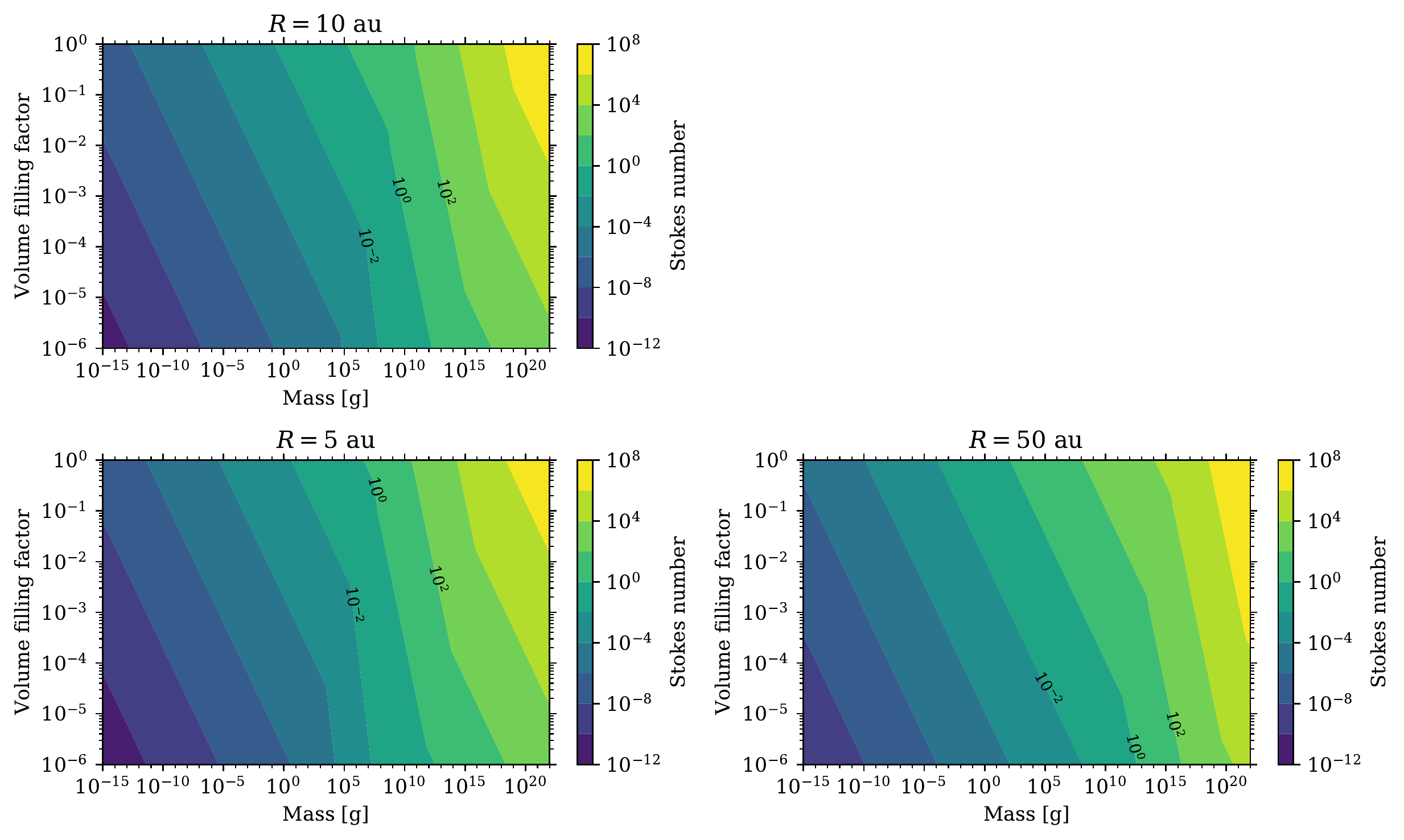}
\caption{{\tatsuuma{Stokes number (color) when $R=10$ au (upper left), $R=5$ au (lower left), and $R=50$ au (lower right).}}
\label{fig:Stall}}
\end{figure}

{\tatsuuma{In this section, we show the Stokes number for every mass and volume filling factor of dust aggregates for every orbital radius in this paper, which we plot in Figure \ref{fig:Stall}.}}
{\fourthcomment{The solid line demarking the boundary of the grey area (the rotational-disruption area in Figures \ref{fig:fiducial}, \ref{fig:forcetorque}, \ref{fig:orbitradius}, and \ref{fig:monomerradius}) is not a line of a constant Stokes number.
However, the Stokes number at the top-left ``cuspy'' corner of the grey area (corresponding to $m\sim10^9\mathrm{\ g}$ and $\phi\sim2\times10^{-2}$ in Figure \ref{fig:fiducial}) is equal to unity.}}
{\tatsuuma{The reason is as follows.
In Figure \ref{fig:tensile}, the bend of the gas-flow torque corresponds to $\St\sim1$ and the tensile stress is constant when $\St>1$.}}
{\fourthcomment{This leads to the result that the Stokes number at the top-left ``cuspy'' corner of the grey area in Figures \ref{fig:fiducial}, \ref{fig:forcetorque}, \ref{fig:orbitradius}, and \ref{fig:monomerradius}  is equal to unity.
However, $\phi$-dependence of the tensile stress due to the gas flow (Equation (\ref{eq:stressdepend}) $\propto \phi^{7/3}$ or $\propto \phi$) does not correspond to that of the tensile strength (Equation (\ref{eq:tensilestrength}) $\propto \phi^{1.8}$).
Therefore, the solid line demarking the boundary of the rotational-disruption area in Figures \ref{fig:fiducial}, \ref{fig:forcetorque}, \ref{fig:orbitradius}, and \ref{fig:monomerradius} is not a line of a constant Stokes number.}}

\bibliography{paper}

\end{document}